\documentclass[12pt]{iopart}
\pdfoutput=1

\usepackage{iopams}  
\usepackage{graphicx}
\usepackage{tabularx}
\usepackage{amssymb}
\usepackage{longtable}
\usepackage{color} 
\usepackage{hyperref}

\DeclareMathAlphabet{\mathitbf}{T1}{cmr}{bx}{it}

\begin{document}

\title[The three dimensional Ising spin glass...]{The three dimensional Ising spin glass in an external magnetic field: the role of the silent majority}

\author{
M. Baity-Jesi$^{1,2,3}$,
R. A. Ba\~nos$^{3,4}$,
A. Cruz$^{3,4}$,
L.A. Fernandez$^{1,3}$,
J.M. Gil-Narvion$^{3}$, 
A. Gordillo-Guerrero$^{3,5}$,
D. I\~niguez$^{3,6}$,
A. Maiorano$^{2,3}$,
F. Mantovani$^{7}$,
E. Marinari$^{8}$,
V. Martin-Mayor$^{1,3}$,
J. Monforte-Garcia$^{3,4}$,
A. Mu\~noz Sudupe$^{1}$,
D. Navarro$^{9}$,
G. Parisi$^{8}$,
S. Perez-Gaviro$^{3}$,
M. Pivanti$^{7}$,
F. Ricci-Tersenghi$^{8}$,
J.J. Ruiz-Lorenzo$^{10,3}$,
S.F. Schifano$^{11}$,
B. Seoane$^{2,3}$,
A. Tarancon$^{3,4}$,
R. Tripiccione$^{7}$ 
and 
D. Yllanes$^{2,3}$}

\address{$^1$  Departamento de F\'\i{}sica Te\'orica I, Universidad Complutense, 28040 Madrid, Spain.} 
\address{$^2$  Dipartimento di Fisica, La Sapienza Universit\`a di Roma, 00185 Roma,  Italy.}
\address{$^3$  Instituto de Biocomputaci\'on y F\'{\i}sica de Sistemas Complejos (BIFI), 50009 Zaragoza, Spain.}
\address{$^4$  Departamento de F\'\i{}sica Te\'orica, Universidad de Zaragoza, 50009 Zaragoza, Spain.} 
\address{$^5$  D. de  Ingenier\'{\i}a El\'ectrica, Electr\'onica y Autom\'atica, U. de Extremadura, 10071, C\'aceres, Spain.}
\address{$^6$  Fundaci\'on ARAID, Diputaci\'on General de Arag\'on, Zaragoza, Spain.}
\address{$^7$  Barcelona Supercomputing Center, 08034 Barcelona, Spain.}
\address{$^8$  Dipartimento di Fisica, IPCF-CNR, UOS Roma Kerberos and INFN, La Sapienza Universit\`a di Roma, 00185 Roma,  Italy.} 
\address{$^9$  D.  de Ingenier\'{\i}a, Electr\'onica y Comunicaciones and I3A, U. de Zaragoza, 50018 Zaragoza, Spain.}
\address{$^{10}$ Departamento de F\'{\i}sica, Universidad de Extremadura, 06071 Badajoz, Spain.}
\address{$^{11}$ Dipartimento di Matematica e Informatica, Universit\`a di Ferrara and INFN, Ferrara, Italy.}

\begin{abstract}
We perform equilibrium parallel-tempering simulations of the 3D Ising
Edwards-Anderson spin glass in a field, using the \emph{Janus} computer. 
A traditional analysis shows
no signs of a phase transition. Yet, we encounter dramatic
fluctuations in the behaviour of the model: Averages over all the data
only describe the behaviour of a small fraction of it. Therefore we
develop a new approach to study the equilibrium behaviour of the
system, by classifying the measurements as a function of a
conditioning variate.  We propose a finite-size scaling analysis based
on the probability distribution function of the conditioning variate,
which may accelerate the convergence to the thermodynamic limit.  In
this way, we find a non-trivial spectrum of behaviours, where a part of
the measurements behaves as the average, while the majority of them
shows signs of scale invariance. As a result, we can estimate the
temperature interval where the phase transition in a field ought to
lie, if it exists. Although this would-be critical regime is
unreachable with present resources, the numerical challenge is
finally well posed.
\end{abstract}

\maketitle 
\tableofcontents
\newpage
\section{Introduction}
Spin glasses are disordered magnetic alloys \cite{mydosh:93}. The most
popular model for these alloys is the Edwards-Anderson
model~\cite{edwards:75,edwards:76}, widely regarded as one of the
simplest instances of a complex system \cite{mezard:87}. The
corresponding phase diagram is tridimensional: Besides the temperature
and the externally applied magnetic field, also the space
dimensionality plays a crucial role \cite{mezard:87}.

Above the upper critical space dimension, $D_\mathrm{u}=6$, mean-field
theory becomes quantitatively accurate
\cite{mezard:87,dealmeida:78,marinari:00}.  As for most magnetic
systems, time-reversal symmetry is spontaneously broken below a
critical temperature $T_\mathrm{c}$. Yet, the behaviour of mean-field
spin glasses in an external field $h$ is most peculiar. 
\footnote{The definition of $h$ is given in section~\ref{sec:model-simulations}, (\ref{eq:ham}).}
Although the
external field explicitly breaks time-reversal symmetry at any
temperature, de Almeida and Thouless have shown that a phase
transition occurs also for $h>0$ at the so-called dAT line,
$T_\mathrm{c}(h)$ \cite{dealmeida:78}. The symmetry that is
spontaneously broken at $T_\mathrm{c}(h)$ is the abstract replica
symmetry, which encodes a complex free-energy landscape
\cite{mezard:87,parisi:79,parisi:80,parisi:83}.

It is still unclear how much of the above picture (usually known as
Replica Symmetry Breaking or RSB picture) is realised in our
three-dimensional world. Some believe it should still apply without
dramatic modifications~\cite{marinari:00,parisi:12}. Yet, a dissenting school of
thought, the so-called droplet picture, predicts no phase transition
in a field as soon as one goes below six spatial dimensions
\cite{mcmillan:84,bray:87,fisher:86,fisher:88b}. Some recent developments of this
debate are in \cite{parisi:12,moore:11,yeo:12}.

A rather obvious way out would be the experimental study of spin
glasses in a field. Unfortunately, opposing indications have been
gleaned over the existence of a phase transition
\cite{jonsson:05,petit:99,petit:02,tabata:10}.

The Renormalisation Group approach to this problem also provides
conflicting results. No fixed points were found by enforcing that the
number of replicas of the replicated field theory be
zero~\cite{bray:80b}. However, fixed points were found relaxing this
condition and using the most general
Hamiltonian~\cite{temesvari:02}. Reasoning along this line, in
\cite{temesvari:08} (see also \cite{parisi:12}) the de
Almeida-Thouless line was computed for $D$ slightly below
$D_\mathrm{u}=6$ (the upper critical dimension remains 6 when
an external magnetic field is applied).

Equilibrium numerical simulations offer an alternative approach, which
has already been effective in establishing that a phase transition
does occur at zero field in the $D=3$ Edwards-Anderson model
\cite{palassini:99,ballesteros:00} (in agreement with
experiments~\cite{gunnarsson:91}). The same strategy has been followed
for $h>0$, with negative results \cite{young:04,jorg:08b}. Yet, this
cannot be the whole story: Recent work in $D=4$, hence below
$D_\mathrm{u}$, using a non-standard finite-size scaling method has
found clear evidence for a dAT line \cite{janus:12}. Furthermore, one
may try to interpolate between $D=3$ and $D=4$ by tuning long-range
interactions in $D=1$ chains \cite{kotliar:85, leuzzi:08}. This
approach suggests that a dAT might be present in $D=4$, but not in
$D=3$ \cite{larson:13} (yet, see the criticism in \cite{leuzzi:13}).

The problem being still open, in \cite{janus:14b} we undertook a
dynamical study of the 3-dimensional Edwards-Anderson spin glass with
the \emph{Janus} dedicated computer, an FPGA-based machine especially
designed for Ising spin glass simulations
\cite{janus:09,janus:10,janus:11,janus:12b}. 
We studied very large lattices ($L=80$), in wide time scales 
(from an equivalent of $\sim1$~ps to $\sim0.01$~s), and gathered 
both equilibrium and non-equilibrium data. We focused on the increase 
of relaxation times and found a would-be dynamical transition, but at 
a suspiciously high temperature. A subsequent examination of the correlation 
length found a growth faster than predicted by the droplet theory, and 
slower than what RSB would expect. We also examined the problem from a 
supercooled liquid point of view 
\cite{debenedetti:97,debenedetti:01,cavagna:09,castellani:05,kirkpatrick:87,kirkpatrick:89}, 
motivated by \cite{moore:02,fullerton:13}. At any rate, the study 
of the possible critical divergence of the correlation length allowed 
us to give upper bounds $T^\mathrm{up}(h)$ to the possible transition 
line for the studied fields.\footnote{In \cite{janus:14b} we studied a bimodal field, while
in this paper $h$ is constant. Notwithstanding, we will make
comparisons with the bounds $T^\mathrm{up}(h)$ by matching
$\overline{h^2}$ in both models.} For further reference we recall
that $T^\mathrm{up}(h=0.1)=0.8$ and
$T^\mathrm{up}(h=0.2)=T^\mathrm{up}(h=0.3)=0.5$.

The impossibility to get concluding evidence in \cite{janus:14b},
may be due to the fact that we did not reach low enough temperatures
(our simulations fell out of equilibrium at temperatures $T$
significantly higher than $T^\mathrm{up}(h)$). In any case, a study of
the equilibrium properties of the model is mandatory if one wants to
understand the nature of the thermodynamic phases of the
three-dimensional Edwards-Anderson spin glass in a field.

In this paper we report the result of equilibrium simulations performed on
\emph{Janus}, using lattices up to $L=32$. Analogously to what has been
already found in mean-field spin glasses on the de Almeida-Thouless line, we
find extreme fluctuations in the model's behaviour \cite{parisi:12b}. We will
propose a method to tame these fluctuations, and we will find out that,
although the average behaviour does not show any sign of a phase transition,
this is not true for the medians of our observables, where we have indications
of a possible phase transition at a temperature $T_\mathrm{c}\lesssim	
T^\mathrm{up}(h)$.

In section~\ref{sec:extended-abstract} we make an extension of this
introduction, giving an intuitive justification of why we deepened our
analysis after finding no sign of a phase transition using the
standard indicators.  We define the model and explain how simulations
were done in section~\ref{sec:model-simulations}. In section~\ref{sec:obs}
we define the observables we measured, and we make a brief parenthesis
on finite-size scaling in
section~\ref{sec:FSS}. Sections \ref{sec:conditional} and \ref{sec:median}
are dedicated to the description of the method we adopted to monitor
extreme fluctuations. In section~\ref{sec:results} we present our
results, and in section~\ref{sec:zero-field} we show that they are not an
echo of the zero-field transition. Finally we give our conclusions in
section~\ref{sec:conclusions}. Further details are given in the
appendices.

\section{Extended introduction: giant fluctuations and the silent minority}
\label{sec:extended-abstract}

\subsection{Foreword}
Since the bulk of the article will get technical in the description of
a method that allows a classification of the data, we
will first give a qualitative description of our results in terms of intuitive
concepts. In this way, the reader will be more aware of the reasons that lead to
the conception of the following analysis.

\subsection{No signs of a phase transition with common tools}
The most common way to locate a phase transition is to identify some observable
$R$ that benefits from scale invariance in the presence of a phase transition (for
example the correlation length $\xi_L$ measured in a lattice of size $L$,
divided by $L$).
This means that if we
plot $R$ as a function of the temperature, for many system sizes, all the
curves will cross at the critical temperature $T_\mathrm{c}$ where the phase
transition occurs. For sufficiently large systems, if the curves do not cross,
there is no phase transition in the simulated temperature range 
(see sections \ref{sec:obs} and \ref{sec:FSS} for details).
The idea dates back to Nightingale \cite{nightingale:76} and Binder \cite{binder:82},
and has been very successful in the study of disordered systems \cite{ballesteros:00,jorg:08b,ballesteros:98,
janus:13b, baityjesi:14,lee:03,jorg:06,leuzzi:08,fernandez:09,
campos:06,banos:12}.

\begin{figure}[t]
\centering
\includegraphics[width=\columnwidth]{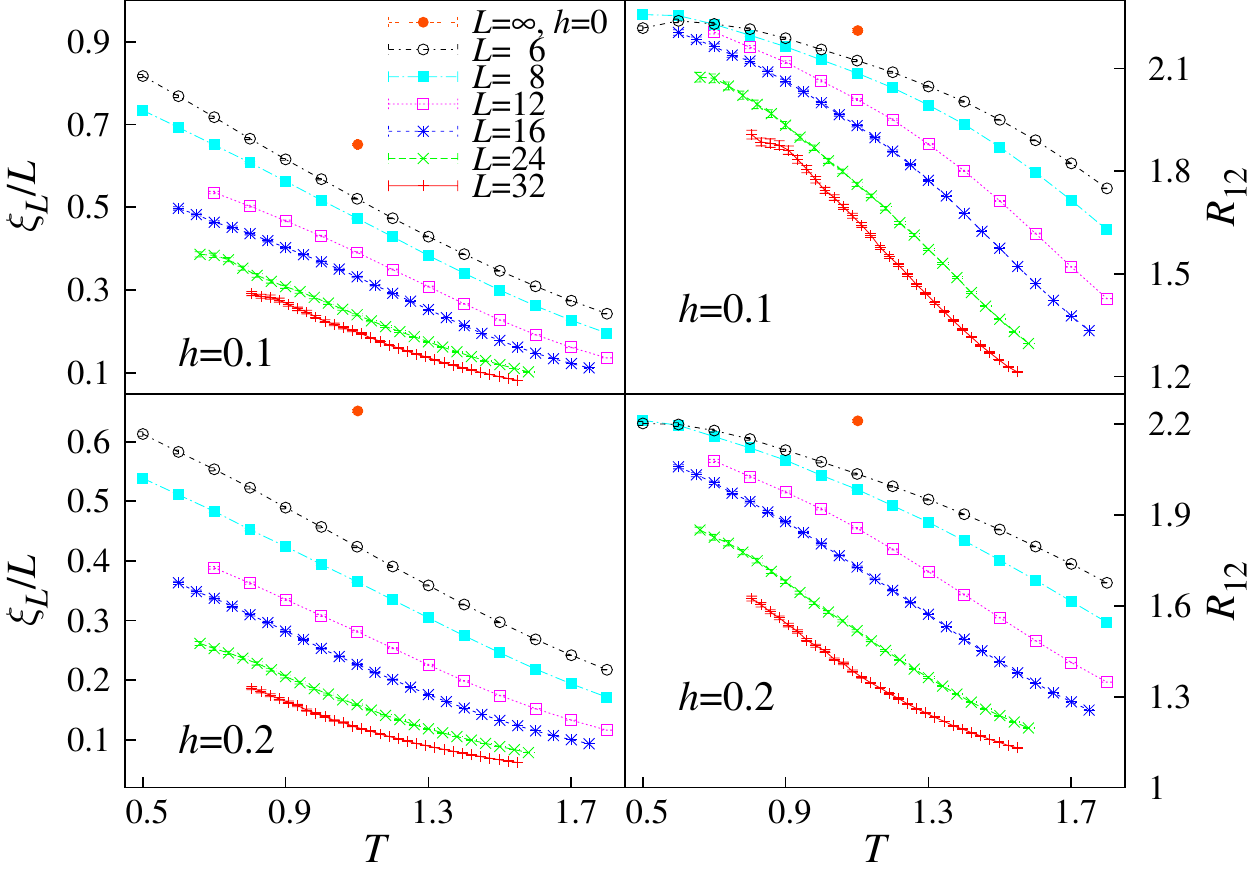}
\caption{The figures on the left show the standard correlation length $\xi_L$ in units of the
  lattice size $L$ as a function of the temperature $T$, for all our lattice
  sizes.  The magnetic fields are $h=0.1$ (top), and $h=0.2$ (bottom). If
  the lattices are large enough, in the presence of a second-order phase
  transition, the curves are expected to cross at a finite temperature
  $T_\mathrm{c}(h)$. 
  The figures on the right show the cumulant $R_{12}$, which in the presence of
  a magnetic field is a better indicator of a phase transition \cite{janus:12},
  for the same magnetic fields.
  At zero field the heights of the crossings (which are universal
  quantities) are indicated
  with a point at $T_\mathrm{c}=1.1019(29)$.
  They are $\xi_L/L(h=0;T_\mathrm{c})=0.6516(32)$ and $R_{12}(h=0;T_\mathrm{c})=2.211(6)$ \cite{janus:13b}.
  In neither case we observe signs of a crossing at the simulated
  temperatures, nor can we state that the curves will cross at lower
  temperature.
  The reader might remark that the curve for $L=32$, $h=0.1$ is not as smooth as
  one would expect from parallel tempering simulations. The reason is twofold.
  On one side the number of simulated samples is much smaller than for $L<32$, 
  and on the other side temperature chaos, which is stronger
  the larger the lattice, is probably present \cite{fernandez:13}.}
\label{fig:xiL}
\end{figure}

In our case, this type of analysis yields a clear result: there is no evidence
of a crossing at the simulated temperatures, magnetic fields and sizes. This is
clearly visible from figure \ref{fig:xiL} where the curves $\xi_L(T)/L$ and $R_{12}(T)$ 
[see (\ref{eq:xi}) and (\ref{eq:R12}) in section~\ref{sec:obs}] 
should have some crossing point if we were in the presence of a phase transition. 
This is in complete qualitative agreement with earlier works on this model \cite{young:04,jorg:08b}.

\subsection{Signs of a hidden behaviour}
Although $\xi_L(T)/L$ is smaller the larger the lattice size, the
coherence length $\xi_L$ grows significantly even for our largest
lattice sizes. For example at $h=0.2$, $T=0.81$ we have
$\xi_{16}=6.09(4)$, $\xi_{24}=7.63(9)$ and $\xi_{32}=9.0(2)$.  The
noticeable size evolution implies that the asymptotic correlation
length $\xi_\infty$ is large compared with $L=32$.

Also, we can examine the behaviour of the spin-glass order parameter,
the overlap $q$, by studying its distribution function $P(q)$.
In the absence of a phase
transition we would be in the paramagnetic phase, and $P(q)$ should be a
delta function of a positive overlap $q_\mathrm{EA}$ (so in finite systems it should
be Gaussian). 

\begin{figure}[t]
 \includegraphics[width=\columnwidth]{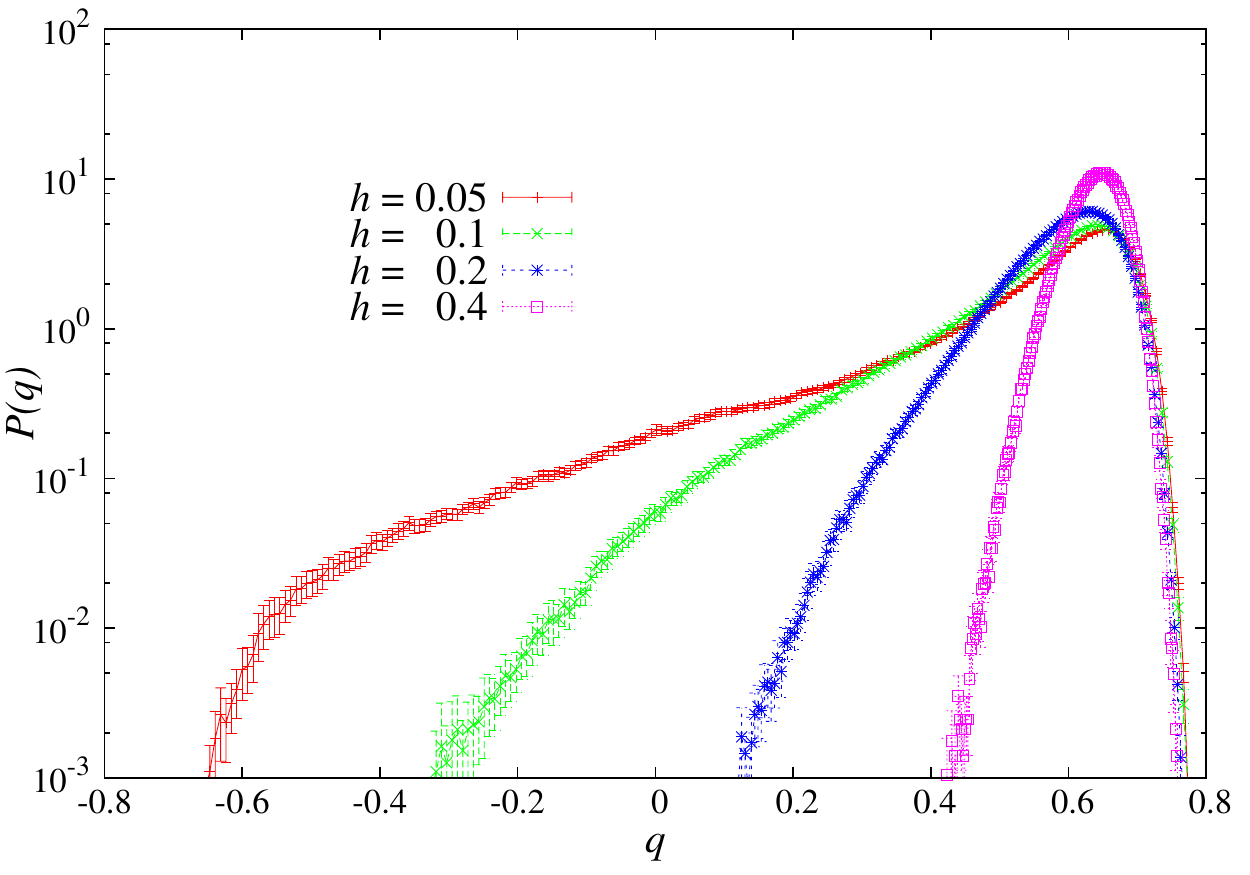}
\caption{The probability distribution function $P(q)$ of the overlap
  $q$, for our largest lattices ($L=32$) at the lowest simulated
  temperature ($T=0.805128$), for all our magnetic fields
  ($h=0.05,0.1,0.2,0.4$), see table~\ref{tab:nsamples}.  The order
  parameter in the Edwards-Anderson model is the overlap $q$, and it
  is defined in the $[-1,1]$ interval (see section~\ref{sec:obs}).  The
  supports are wide, with exponential tails similar to those in the
  mean-field model at the dAT transition line
  \cite{parisi:12b}.}
\label{fig:Pq}
\end{figure}

Instead, we can see from figure~\ref{fig:Pq} that its distribution
$P(q)$ has a very wide support, with tails that, for small enough
magnetic fields, reach even negative values of $q$. This is precisely
what was observed in the mean-field version of the model on the de
Almeida-Thouless line, and it was attributed to the contribution of
few samples \cite{parisi:12b}.

From these arguments it becomes reasonable to think that we may not be simulating large enough
lattices to observe the asymptotic nature of the system and that there may be
some hidden behaviour that we are not appreciating.

\subsection{Giant fluctuations}
In fact, we find out that the average values we measure are
representative  of only a small part of the data set.  That is, the
average of relevant observables (e.g., the spatial correlation
function) only represents the small number of measurements that are
dominating it.  The rest of the measurements is not appreciated by using
the average.

\begin{figure}[t]
\includegraphics[width=\columnwidth]{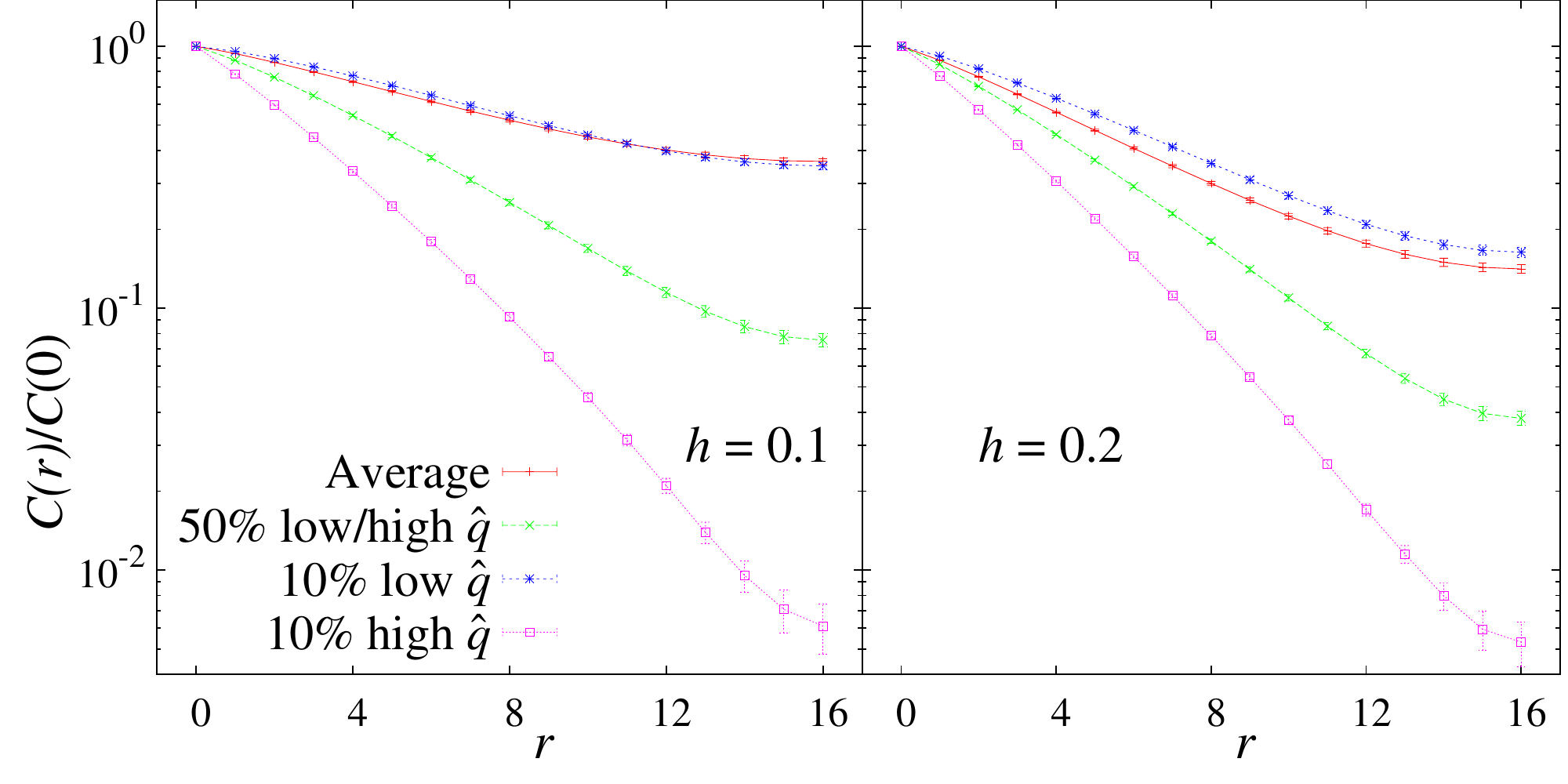}
\caption{Different instances of the normalised correlation function $C(r)$ (\ref{eq:Crplane}) for $L=32$, $T=0.805128$.
The field is $h=0.1$ on the left, and $h=0.2$ in the right plot.
We sort the measurements with the help of a conditioning variate $\hat q$ as described in section~\ref{sec:conditional}. 
In this case $\hat{q}$ is the median overlap $q_\mathrm{med}$.
We show small sets of measurements. Namely, the ones with the $10\%$ lowest (top curve) and
highest (bottom curve) $\hat{q}$ and those whose $\hat q$ corresponds to the median
of the distribution of $\hat q$ ($50\%$ lowest/highest $\hat{q}$).
This sorting reveals extreme differences in
the \emph{fauna} of measurements.
The average and median of the correlation
functions are very different. The average is very similar to the $10\%$ lowest
ranked measures, i.e., it is only representative of a very small part of the
data. 
We normalise $C(r)$ by dividing by $C(0)$ because we measure
point-to-plane correlation functions (\ref{eq:Crplane}).
The correlation functions have zero slope at $r=L/2$ due to the
periodic boundary conditions.}
\label{fig:Cr}
\end{figure}

Clearly, standard finite-size scaling methods are not adequate to these 
systems, and we need to find a way to take into account \emph{all} the 
measurements. Recalling the wide distributions of figure~\ref{fig:Pq}, 
it seems reasonable to sort our measurements according to some conditioning 
variable $\hat q$ related to the overlaps between our replicas 
(see section \ref{sec:conditional}). This way, we find out that the
average values we measure are given by only a small part of the measurements. For
example in figure \ref{fig:Cr} we show the correlation function $C(r)$. We plot 4
estimators of $C(r)$: the average (which is the standard quantity studied in almost
all, if not all, previous work), the $C(r)$ that corresponds 
to the median of the $\hat q$ distribution, 
and the measurements with the $10\%$ highest (lowest) value
of $\hat{q}$. We see that the
average is very close to the $10\%$ lowest $\hat q$, and very far from the two other
curves. So, when we plot the average curve, we are only representing the
behaviour of that small set of data. 

Therefore, if we want to understand the behaviour of the \emph{whole} collection of measurements, we
have to be able to find some criterion to sort them and analyse them separately.

\section{Model and simulations}
\label{sec:model-simulations}
\subsection{The model}
We consider a 3$D$ cubic lattice of size $L$ with periodic boundary conditions. 
In each of the $V=L^3$ vertices of the lattice there is a spin 
$\sigma_\mathitbf{x} = \pm1$. The spins interact uniquely with their nearest
neighbours and with an external magnetic field $h$.  The Hamiltonian is
\begin{equation}
  \label{eq:ham}
  {\cal H} = - \sum_{\langle \mathitbf{x},\mathitbf{y} \rangle}J_{\mathitbf{x y}} \sigma_\mathitbf{x} \sigma_\mathitbf{y} 
              - h\sum_\mathitbf{x} \sigma_\mathitbf{x}\,,
\end{equation}
where $\langle \mathitbf{x},\mathitbf{y} \rangle$ means the sums are
only over the nearest neighbours, while the couplings $J_{\mathitbf{x
    y}}$, which are constant during each simulation, take the values
$\pm1$ with equal probability (quenched disorder). A given instance of
the bonds $J_{\mathitbf{x y}}$ and of the intensity of the magnetic
field $h$ define a \emph{sample}.  We will consider real
\emph{replicas} of each sample, i.e., systems with identical couplings
$J_\mathitbf{x y}$ and field $h$, but independent evolutions (for a
recent discussion see \cite{janus:10} and \cite{janus:09b}). In
this work we will use 4 replicas per sample.

\subsection{The simulations}
For all our simulations we made use of parallel tempering
(PT) \cite{Hukushima:96,Marinari:98b}. The whole procedure was very similar to
the one in \cite{janus:12}.

The smaller lattices ($L=6,8,12$) were simulated 
with multispin coding (C code with words of 128 bits, by means of streaming
extensions) \cite{janus:12,newman:99} on the \emph{Memento} CPU cluster 
at BIFI.
The larger samples ($L=16,24,32$) were
simulated on the \emph{Janus} dedicated computer \cite{janus:12b}.  

An Elementary Monte Carlo
Step (EMCS) consisted in 1 PT exchange every 10 Metropolis steps for the multispin-coding
samples, and 1 PT every 10 heatbath for the samples simulated on \emph{Janus}.
table \ref{tab:nsamples} shows the relevant parameters of the simulations. The
temperatures were equally spaced between $T_\mathrm{min}$ and $T_\mathrm{max}$.
The intensities of the external magnetic field we chose are $h=0.05,0.1,0.2$
and $0.4$.

To check whether the samples were thermalised we measured the exponential
self-correlation time of the PT random walk in temperatures
$\tau$ \cite{janus:12,janus:10,fernandez:09b,yllanes:11}.
 We required the simulations to last at least
$14\tau$. 
To do so without consuming computing time on already thermalised
lattices, we assigned a minimum number of EMCS, $N_\mathrm{MCS}^{\mathrm{min}}$,
for all the samples, and extended by a factor $f>1$ only the ones that did not
meet the imposed thermalisation criterion.  In table \ref{tab:nsamples} we
report $N_\mathrm{EMCS}^{\mathrm{min}}$, the maximum extension factor
$f_\mathrm{max}$ of the simulations, and minimum number $N^\mathrm{min}_\tau$
of EMCS in units of $\tau$.

Equilibrium measures were taken offline over the second half of each
simulation.  Independently of how much the simulations were extended,
we saved $16$ equally time-spaced configurations and performed
measures on them. We were measuring 4-replica observables. Therefore,
for each sample it was possible to choose quadruplets of
configurations, each from a different replica, in $16^4$ ways.  Out of
the $16^4$ possibilities, we chose randomly $N_\mathrm{t} = 1000$
combinations.  In other words, each sample participated in the
statistics with $N_\mathrm{t}=1000$ measurements.

The errors were estimated with the jackknife method.

\begin{table}[htb]
 \caption{Parameters of the simulations. We report the magnetic field
   $h$, the lattice linear size $L$, the number of simulated samples
   $N_\mathrm{samples}$, and the basic length of a simulation in
   Elementary Monte Carlo Steps $N_\mathrm{EMCS}^{\mathrm{min}}$.  In
   each simulation we measured the exponential correlation time $\tau$
   of the PT random walk in temperatures.  When $\tau$ was too large
   to meet our thermalisation requirements, we extended the length of
   each simulation by an extension factor $f$.  We denote with
   $f_\mathrm{max}$ the greatest extension factor.  We also give the
   minimum length of a simulation $N^\mathrm{min}_\tau$ in units of
   $\tau$.  In all cases we imposed $N^\mathrm{min}_\tau>14$.
   Finally, we give the number of temperatures
   $N_T$ we used for the PT, and the minimum and maximum temperatures
   $T_\mathrm{min}$ and $T_\mathrm{max}$.  }
\label{tab:nsamples}
\begin{tabular*}{\columnwidth}{@{\extracolsep{\fill}}ccccccccc}
$h$ & $L$ & $N_\mathrm{samples}$ & $N_\mathrm{EMCS}^{\mathrm{min}}$& $f_\mathrm{max}$ & $N^\mathrm{min}_\tau$ &$N_\mathrm{T}$  & $T_\mathrm{min}$ & $T_\mathrm{max}$\\\hline\hline
0.05  &  6  & 25600 &  $1.6\times10^6$ &   1 & 40.0 & 14 & 0.5 & 1.8 \\
0.05  &  8  & 25600 &  $3.2\times10^6$ &  16 & 40.0 & 14 & 0.5 & 1.8 \\
0.05  & 12  & 25600 &  $3.2\times10^6$ &  16 & 15.6 & 12 & 0.7 & 1.8 \\
0.05  & 16  & 12800 & $1.28\times10^7$ & 128 & 20.1 & 24 & 0.6 & 1.75 \\
0.05  & 24  &  6400 & $1.28\times10^7$ & 110 & 16.0 & 20 & 0.78 & 1.54 \\
0.05  & 32  &  2400 & $6.4\times10^7$ & 256 & 14.3 & 30 & 0.805128 & 1.54872 \\\hline
0.1   &  6  & 25600 &  $1.6\times10^6$ &   4 & 40.0 & 14 & 0.5 & 1.8 \\
0.1   &  8  & 25600 &  $3.2\times10^6$ &  16 & 40.0 & 14 & 0.5 & 1.8 \\
0.1   & 12  & 25600 &  $3.2\times10^6$ &  16 & 14.4 & 12 & 0.7 & 1.8 \\
0.1   & 16  & 12800 & $1.28\times10^7$ & 256 & 27.9 & 24 & 0.6 & 1.75 \\
0.1   & 24  &  3200 & $1.28\times10^7$ &4097 & 14.3 & 24 & 0.66 & 1.58 \\
0.1   & 32  &  1600 & $6.4\times10^7$ & 533   & 14.4    & 30 & 0.805128 & 1.54872 \\\hline
0.2   &  6  & 25600 &  $1.6\times10^6$ &   1 & 40.0 & 14 & 0.5 & 1.8 \\
0.2   &  8  & 25600 &  $3.2\times10^6$ &  16 & 40.0 & 14 & 0.5 & 1.8 \\
0.2   & 12  & 25600 &  $3.2\times10^6$ &  64 & 25.4 & 12 & 0.7 & 1.8 \\
0.2   & 16  & 12800 & $1.28\times10^7$ & 256 & 18.4 & 24 & 0.6 & 1.75 \\
0.2   & 24  &  3200 & $1.28\times10^7$ & 512 & 16.1 & 24 & 0.66 & 1.58 \\
0.2   & 32  &  1600 & $1.6\times10^7$ & 513 & 16.0 & 30 & 0.805128 & 1.54872 \\\hline
0.4   &  6  & 25600 &  $1.6\times10^6$ &   1 & 40.0 & 14 & 0.5 & 1.8 \\
0.4   &  8  & 25600 &  $3.2\times10^6$ &   4 & 30.7 & 14 & 0.5 & 1.8 \\
0.4   & 12  & 25600 &  $3.2\times10^6$ &  16 & 14.1 & 12 & 0.7 & 1.8 \\
0.4   & 16  &  3200 & $1.28\times10^7$ &  32 & 20.1 & 24 & 0.6 & 1.75 \\
0.4   & 24  &   800 & $1.28\times10^7$ &  29 & 16.1 & 24 & 0.66 & 1.58 \\
0.4   & 32  &   800 &  $3.2\times10^6$ &  16 & 16.4 & 30 & 0.805128 & 1.54872 \\\hline
\end{tabular*}
\end{table}

\section{Observables}
\label{sec:obs}
For each sample we simulated 4 different replicas, in order to be able to
compute connected correlation functions that go to zero at infinite distance. 
We will
label replicas by using superscripts, this means that the generic quantity ${\cal
O}^{(a)}$ belongs to the $a^\mathrm{th}$ replica of a given sample.

We will denote with over-lines $\overline{(\ldots)}$ the averages over the
samples, and with brackets $\langle\ldots\rangle$ the thermal averages. 
To make notation less heavy we will use $E(\ldots)=\overline{\langle\ldots\rangle}$ 
to denote an average that is taken first over the
thermal fluctuations and then over the samples.

We define the local overlap as
\begin{equation}
 \label{eq:qx}
 q^\mathrm{(ab)}_\mathitbf{x}=\sigma_\mathitbf{x}^{(a)} \sigma_\mathitbf{x}^{(b)}\,,
\end{equation}
while the total overlap is
\begin{equation}
\label{eq:q}
 q^{(ab)}=\frac{1}{V} \sum_\mathitbf{x} q^{(ab)}_\mathitbf{x}\,.
 \end{equation}

We show in~\ref{app:GL} that the most informative connected
correlator we can construct with 4 replicas is the replicon propagator \cite{dealmeida:78,dedominicis:06}
\begin{equation}
\label{eq:GR}
G_\mathrm{R}(\mathitbf{r})=\frac{1}{V} \sum_{\mathitbf{x}} 
  \overline{\left( \langle \sigma_{\mathitbf{x}} \sigma_{\mathitbf{x}+\mathitbf{r}} \rangle 
                 - \langle \sigma_{\mathitbf{x}} \rangle \langle \sigma_{\mathitbf{x}+\mathitbf{r}} \rangle 
	    \right)^2}\,.
\end{equation}
To compute $G_\mathrm{R}$ we calculate the 4-replica field
\begin{equation}
 \Phi_{x}^{(ab;cd)}=\frac{1}{2}(\sigma_\mathitbf{x}^{(a)} - \sigma_\mathitbf{x}^{(b)}) (\sigma_\mathitbf{x}^{(c)} - \sigma_\mathitbf{x}^{(d)})\,,
\end{equation}
where the indices $a,b,c,d$ indicate strictly different replicas.
Notice that
\begin{equation}
  \left\langle \Phi_{\mathitbf{x}}^{(ab;cd)} \Phi_{\mathitbf{y}}^{(ab;cd)} \right\rangle
  =
  \left( \left\langle 
    \sigma_{\mathitbf{x}} \sigma_{\mathitbf{x}+\mathitbf{r}} \rangle 
      - \langle \sigma_{\mathitbf{x}} \rangle \langle \sigma_{\mathitbf{x}+\mathitbf{r}} \right\rangle 
	    \right)^2 \,,
\end{equation}
 so we obtain $G_\mathrm{R}$ by taking also the average over the samples
\begin{equation}
 E(\Phi_{\mathitbf{x}}^{(ab;cd)} \Phi_{\mathitbf{y}}^{(ab;cd)}) = G_\mathrm{R}(\mathitbf x - \mathitbf y)\,.
 \label{eq:GRxy}
\end{equation}
Here, and everywhere there is more than one possible permutation of the replica indices,
we average over all of them to gain statistics.

Correlations in the Fourier space are defined analogously, by Fourier-transforming
$\Phi_{\mathitbf{x}}^{(ab;cd)}$, so the wave-vector dependent replicon susceptibility
is expressed as
\begin{equation}
\label{eq:chi}
\chi_\mathrm{R}(\mathitbf{k}) = \frac{1}{V} E(|\hat {\Phi}_{\mathitbf{k}}^{(ab;cd)}|^2)
~~~,~~~
 \hat {\Phi}_{\mathitbf{k}}^{(ab;cd)} = \sum_\mathitbf{x}^V \mathrm{e}^{\rmi\mathitbf{k}\cdot\mathitbf{x}}\Phi_{\mathitbf{x}}^{(ab;cd)}\,.
\end{equation}
When we omit $\mathitbf{k}$, we refer to the susceptibility $\chi = \chi_\mathrm{R}(\mathitbf{0})$. 

We compute point-to-plane correlation functions
\begin{equation}
\label{eq:Crplane}
 C(r) ~=~ \frac{1}{L} \sum_{n=0}^{L-1} \mathrm{e}^{-\rmi r 2\pi n/L} \chi_\mathrm{R}\big(\frac{2\pi n}{L},0,0\big)     
      ~\equiv~ \sum_{y,z} G_\mathrm{R}(x=r,y,z)\,.
\end{equation}
The previous relation is equivalent if we align the wave vector along any of the three coordinate axes,
so we average over these three choices.

With the defined observables we were able to calculate the second-moment correlation length
\begin{equation}
 \xi_L = \frac{1}{2 \sin{(k_\mathrm{min}/2)}} \sqrt{\frac{\chi_\mathrm{R}(\bf{0})}{\chi_\mathrm{R}(2\pi/L,0,0)}-1}\,,
 \label{eq:xi}
\end{equation}
and the dimensionless quantity
\begin{equation}
 R_{12} = \frac{\chi_\mathrm{R}(2\pi/L,0,0)}{\chi_\mathrm{R}(2\pi/L,\pm2\pi/L,0)}\,
 \label{eq:R12}
\end{equation}
where all the quantities were averaged over all the possible permutations of the components of the wave vectors.

The cumulant $R_{12}$ (recall figure \ref{fig:xiL}) was used in
\cite{janus:12} to estimate the critical temperature bypassing pathologies
on $\chi({\bf 0})$ due to the fact that the overlap is non-zero in the
paramagnetic phase (recall figure \ref{fig:Pq}).

\section{Finite-size scaling}
\label{sec:FSS}
Our finite-size scaling analysis follows phenomenological
renormalisation, which is very effective when one looks for a phase
transition with a diverging correlation length
\cite{amit:05,nightingale:76,ballesteros:96}. In fact, under the
accepted assumption that there is only one relevant length scale in
the system, close to the critical temperature $T_\mathrm{c}$ a generic
observable ${\cal O}$ scales like
\begin{equation}
 \left\langle{\cal O}(L,T)\right\rangle = L^{x_{\cal O}/\nu} \left[{\cal F}_{\cal O}(L^{1/\nu}(T-T_\mathrm{c})) + O(L^{-\omega})\right]\,,
\end{equation}
where $\omega>0$ represents the leading irrelevant exponent.

This implies that at the critical point the quantities $\xi_L(T,L,h)/L$ and
$R_{12}(T,L,h)$ are scale invariant to the dominant order:
\begin{eqnarray}
\label{eq:xiL-invariance}
\frac{\xi_L}{L}(L,T) &= {\cal F}_\xi (L^{1/\nu}(T-T_\mathrm{c})) + \ldots\,,\\[2ex]
\label{eq:R12-invariance}
R_{12}(L,T) &= {\cal F}_R (L^{1/\nu}(T-T_\mathrm{c})) + \ldots\,,
\end{eqnarray}
where the dots indicate subleading corrections to scaling.
So, for large enough
systems the critical point at given $h$ would be revealed by the crossing
points of these curves for different sizes. As figure~\ref{fig:xiL} shows,
such a crossing is not present in our data.

With the use of the hyperscaling relations we can also predict the
scaling of the susceptibility close to the critical point:
\begin{equation}
\label{eq:chi-scaling}
 \chi_\mathrm{R} (L,T) = L^{2-\eta} {\cal F}_\chi (L^{1/\nu}(T-T_\mathrm{c})) + \ldots\,.
\end{equation}

\section{Conditional expectation values and variances}
\label{sec:conditional}

\subsection{The conditioning variate}
As we pointed out in section~\ref{sec:extended-abstract}, the behaviour of the system 
is dominated by a very small number of measurements.

This means that the average over all the measurements of an
observable does not describe the typical behaviour of the system. 
Furthermore, the behaviour of the measurements that contribute less to the full 
averages is qualitatively different from the one of those who give the 
main contribution (see figure~\ref{fig:Cr} and later on section~\ref{sec:results}).

We want to classify our measurements in a convenient way, in order to be able
to separate different behaviours, and analyse them separately. To
this goal, we replace normal expectation values $E({\cal{O}})$ of a generic
observable ${\cal O}$, with the expectation value $E({\cal{O}}|\hat{q})$
conditioned to another random variable $\hat q$. 
Perhaps for lack of imagination $\hat q$ will be named conditioning variate.
For each instance of
${\cal O}$ we monitor also the value of $\hat q$, and we use it to label ${\cal O}$. 
Hopefully, there will be some correlation.

The conditional expectation value is defined as the average of ${\cal O}$,
restricted to the measurements $i$ (out of the ${\cal N_\mathrm{m}}=N_\mathrm{t}N_\mathrm{samples}$ total
measurements) that simultaneously yield ${\cal O}_i$ and
$\hat q_i$ [so we are actually talking
about couples of simultaneous measurements $({\cal O}_i,\hat q_i)$]
in a small interval around $\hat q=c$, 
\begin{equation}
 E({\cal O}|\hat q = c)~=~  \frac{E\left[{\cal O}_i {\cal X}_{\hat q = c}(\hat q_i)\right]}{E\left[{\cal X}_{\hat q = c}(\hat q_i)\right]}\,.
\end{equation}
Where we have used the characteristic function
\begin{equation}
  {\cal X}_{c}(\hat q_i) = \left\{
\begin{array}{rl}
1,& ~\mathrm{ if }~ |c - \hat q_i| < \epsilon \sim \frac{1}{\sqrt{V}} \\
0,& ~\mathrm{ otherwise}.
\end{array}
\right.
\end{equation}
In~\ref{app:hack} we give technical details on the choice of
$\epsilon$. To make notation lighter, in the rest of the paper we will replace $E({\cal O}|\hat q = c)$
with $E({\cal O}|\hat q	)$.

The traditional expectation value $E({\cal{O}})$ can be recovered by integrating
over all the possible values of the conditioning variate $\hat{q}$:
\begin{equation}
 \label{eq:E-conditioned}
 E({\cal{O}}) = \int \rmd\hat{q}\ E({\cal O}|\hat{q}) P(\hat{q})~~,~~ P(\hat q) = E[{\cal X}_{\hat q}]
~,
\end{equation}
where $P(\hat q)$ is the probability distribution function of the
conditioning variate.

We remark that the concept of conditioning variate is fairly similar to the one of
control-variate. Yet, the latter was formalised slightly differently, and with
the objective of enhancing the precision of the measures \cite{fernandez:09c}. In
\cite{janus:10,janus:10b} a procedure very similar to the present
one was followed, but the aim was constructing clustering correlation
functions, while in our case the conditioning variate is used to analyse
\emph{separately} different behaviours outcoming from the same global data set,
so that a sensible finite-size scaling becomes possible.

\subsection{Measurements against samples}

\begin{figure}[b]
 \includegraphics[width=\columnwidth]{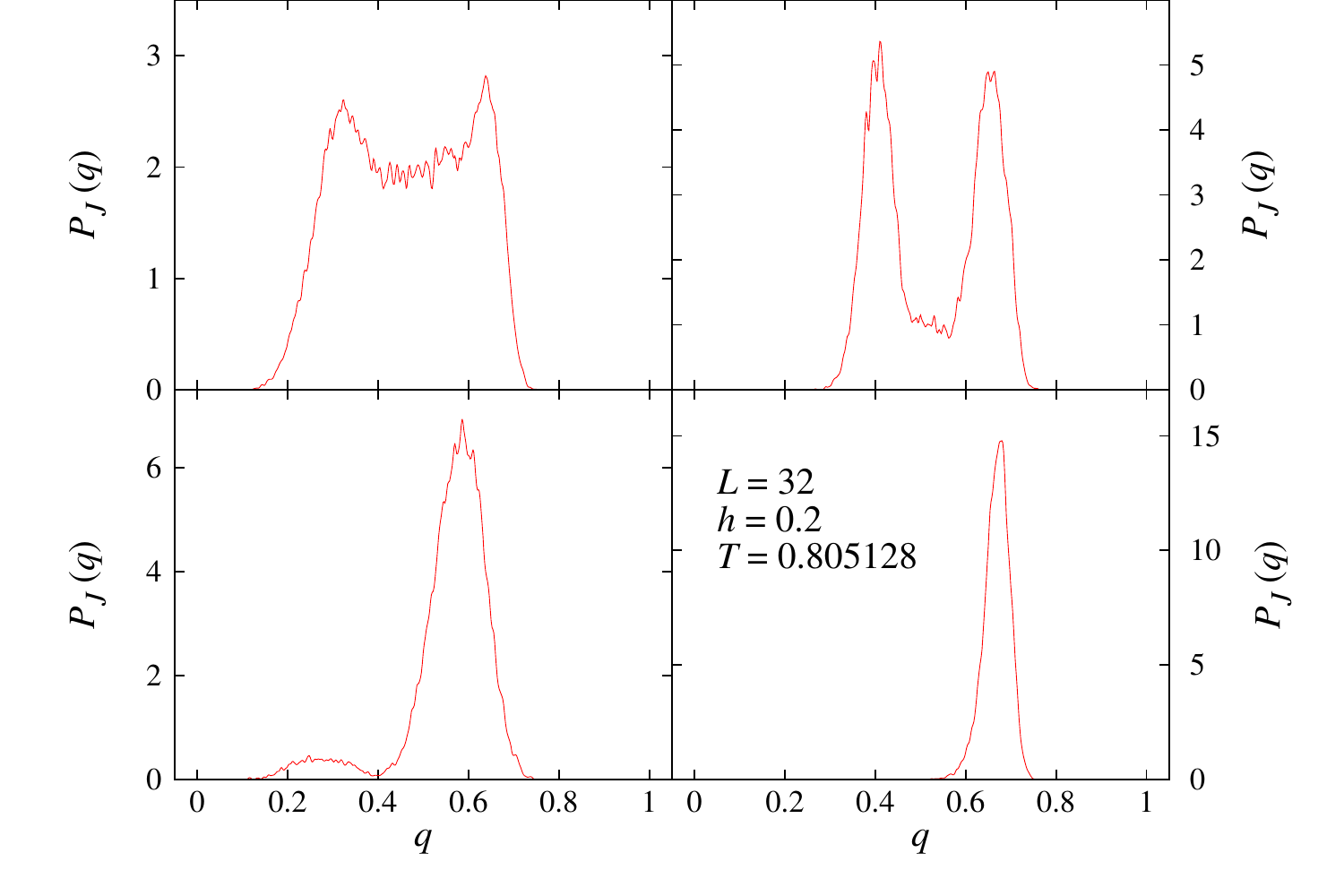}
 \caption{Sample-dependent probability distribution functions $P_J(q)$, for four different samples,
	  each representing a different type of $P_J(q)$ we encountered.
	  As well as the averaged $P(q)$, also the sample-dependent density function can be
	  wide and with a structure. The plotted data comes from samples with $L=32$, $h=0.2$ and $T=0.805128$.}
 \label{fig:PJq}
\end{figure}

The reader may argue that a sample-to-sample distinction of the
different behaviours is more natural than a measurement-dependent one
(although intuition leads to assume that the two are related). This
was indeed our first approach to the problem (it was, in fact,
proposed in Ref.~\cite{parisi:12b}). However, we found that the
approach described in the previous section is preferable,
both for practical and conceptual reasons.

On the practical side, a sample-to-sample separation implies that from
each sample we get only one data point: For any observable,
we limit ourselves to its thermal average.  In this case we would need a
limitless amount of samples to be able to construct a reasonable
$P(\hat q)$. Moreover, the simulations should last a huge number of
autocorrelation times $\tau$ if we want to have small enough errors on
the thermal averages of each sample. Otherwise, we would introduce a
large bias that is not reduced when increasing the number of samples.

On the conceptual side, representing each sample merely with a single
number (namely the thermal expectation value), is a severe
oversimplification. As we show in Fig.~\ref{fig:PJq}, even though we
are in the paramagnetic phase, the behaviour within each sample is far
from trivial. For a non-negligible fraction of the samples, the
overlap distribution is wide, often with a multi-peak structure. The
barriers among peaks can be deep, hence suggesting extremely slow
dynamics (which is indeed the case for physical dinamics,
Ref.~\cite{janus:14b}, or for the parallel tempering dynamics,
section~\ref{sec:model-simulations}). 

In summary, we find that using instantaneous measurements to classify
the available information is the best solution.

\subsection{The selection of the conditioning variate}
\label{subsec:select-cv}
\subsubsection{A quantitative criterion}
A convenient conditioning variate is
the one that mostly discerns the different behaviours of the model.  We can get a
quantitative criterion for the selection of a good $\hat q$ by examining the following
relation for any conditional variance:
\begin{equation}
\label{eq:regla-de-suma-var}
\mathrm{var}({\cal O}) = c_1 + c_2,
\end{equation}
where we defined
\begin{eqnarray}
 \nonumber
c_1 &\equiv& \int_{-1}^1 \rmd\hat{q} \, P(\hat{q})\mathrm{var}({\cal O} | \hat{q}) ~~,~~~~~~\mathrm{var}({\cal O} | \hat{q}) = E([{\cal O} - E({\cal O}|\hat{q})]^2 ~|~ \hat{q}) \,,
\\ 
\label{eq:c1-c2}
c_2 &\equiv& \int_{-1}^1 \rmd\hat{q} \, P(\hat{q})[ E({\cal O}) - E({\cal O}|
  \hat{q})]^2\,.
\end{eqnarray}
Both $c_1$ and $c_2$ are positive, and their sum is fixed.

Let us explain intuitively why a useful conditioning variate has $c_2\gg c_1$.

If $c_1=0$ the fluctuations of ${\cal O}$
would be explained solely by the fluctuations of $\hat q$. 
In this case $c_2$ is large and assume its largest possible value, meaning that
different values of ${\cal O}$ are mostly spread apart by $\hat q$.

On the other side, $c_2=0$ implies $E({\cal O}) = E({\cal O}|\hat q)$ and
signals an insensitive conditioning variate, with null correlation 
between ${\cal O}$ and $\hat{q}$.

Equations (\ref{eq:regla-de-suma-var}) and (\ref{eq:c1-c2}) can thus be used to quantify the quality of the
conditioning variate $\hat{q}$: We look for the highest quotient $c_2/c_1$.

\subsubsection{Candidates for \texorpdfstring{$\hat q$}{q}}

To select an appropriate conditioning variate we need to chose ${\cal O}$
and propose some test definitions for $\hat q$.
The functions of the observables that one could use as a conditioning variate
are infinite, but physical intuition lead us to try with simple functions of
the overlap. 
On the other side, a natural choice of ${\cal O}$ is the estimator of the replicon 
susceptibility [see (\ref{eq:chi})]. This means that
\begin{equation}
 {\cal O} \longrightarrow \frac{1}{3{\cal N}}\sum^{\cal N}_{\stackrel{\mathrm{equiv. wave}}{\mathrm{vectors}~\mathitbf{k}}}\left[\, |\Phi_{\mathitbf{k}}^{(ab;cd)}|^2 
										     + |\Phi_{\mathitbf{k}}^{(ac;bd)}|^2 
										     + |\Phi_{\mathitbf{k}}^{(ad;bc)}|^2 \,\right]\,,
\end{equation}
where ${\cal N}$ is the number of equivalent wave vectors one can construct.
This is a 4-replica quantity [see (\ref{eq:GR})], so 6 instantaneous overlaps are 
associated to each instance of the correlators. To define $\hat q$ we need to
propose a function of the six overlaps in order to get a one-to-one correspondence.

Let us reorder each 6-plet of instantaneous overlaps $\{q^{(ij)}\}$ in the form
of six sorted overlaps $\{q_{k}\}$
\begin{equation}
 \left\{q^{(ab)},q^{(ac)},q^{(ad)},q^{(bc)},q^{(bd)},q^{(cd)}\right\} 
\longrightarrow 
\left\{q_1\leq q_2\leq q_3\leq q_4\leq q_5\leq q_6\right\}\,.
\end{equation}
The following are natural test conditioning variates:
\begin{equation}
\label{eq:cv}
\hat{q} =\left\{
\begin{array}{llr}
q_\mathrm{min}&=q_1&\mathrm{(the~minimum)}\\[1ex]
q_\mathrm{max}&=q_6&\mathrm{(the~maximum)}\\[1ex]
q_\mathrm{med}&=\frac{1}{2}(q_3+q_4)&\mathrm{(the~median)}\\[1ex]
q_\mathrm{av }&=\frac{1}{6}(q_1+q_2+q_3+q_4+q_5+q_6)&\mathrm{(the~average)}\,.\\
\end{array}
\right.
\end{equation}
table~\ref{tab:reglas-de-suma} depicts the $c_1$ and $c_2$ terms, and their
ratio, for all the conditioning variates, for a single triplet $(T,L,h)$ 
and $\mathitbf{k}=(0,0,0)$. The
best conditioning variate is clearly the median, since it has the highest
$c_2/c_1$ ratio. The situation is similar for other choices of $(T,L,h)$.
\begin{table}[tb]
\centering
\caption{Criterion for the choice of the
conditioning variate $\hat q$ for $h=0.1$, $L=32$, $T=0.805128$, by looking at the 
indicators $c_1$ and $c_2$ relatively to $\chi_\mathrm{R}(\bf{0})$. We want the
$\hat q$ to split as much as possible the different measured susceptibilities.
This is obtained, see (\ref{eq:c1-c2}), when the ratio $c_2/c_1$ is
maximised. From the data we see that this occurs with $\hat{q} = 
q_{\mathrm{med}}$. }
\label{tab:reglas-de-suma}
\begin{tabular*}{\columnwidth}{@{\extracolsep{\fill}}cccc}
$\hat{q}$ & $c_1$ & $c_2$ & $c_2/c_1$\\\hline\hline
  $q_\mathrm{min}$ & 399000$\pm$ 37000 & 121000 $\pm$ 15000 &  0.30(6) \\[1ex] 
  $q_\mathrm{max}$ & 514000$\pm$ 51000 & 6230 $\pm$ 690 & 0.012(3)\\[1ex] 
  $q_\mathrm{med}$ & 162000$\pm$ 10000 & 358000$\pm$ 45000 & 2.2(4)\\[1ex]
  $q_\mathrm{av}$ & 328000$\pm$ 26000 & 192000 $\pm$ 28000 & 0.6(1)
\end{tabular*}
\end{table}

\begin{figure}[tbh]
 \includegraphics[width=\columnwidth]{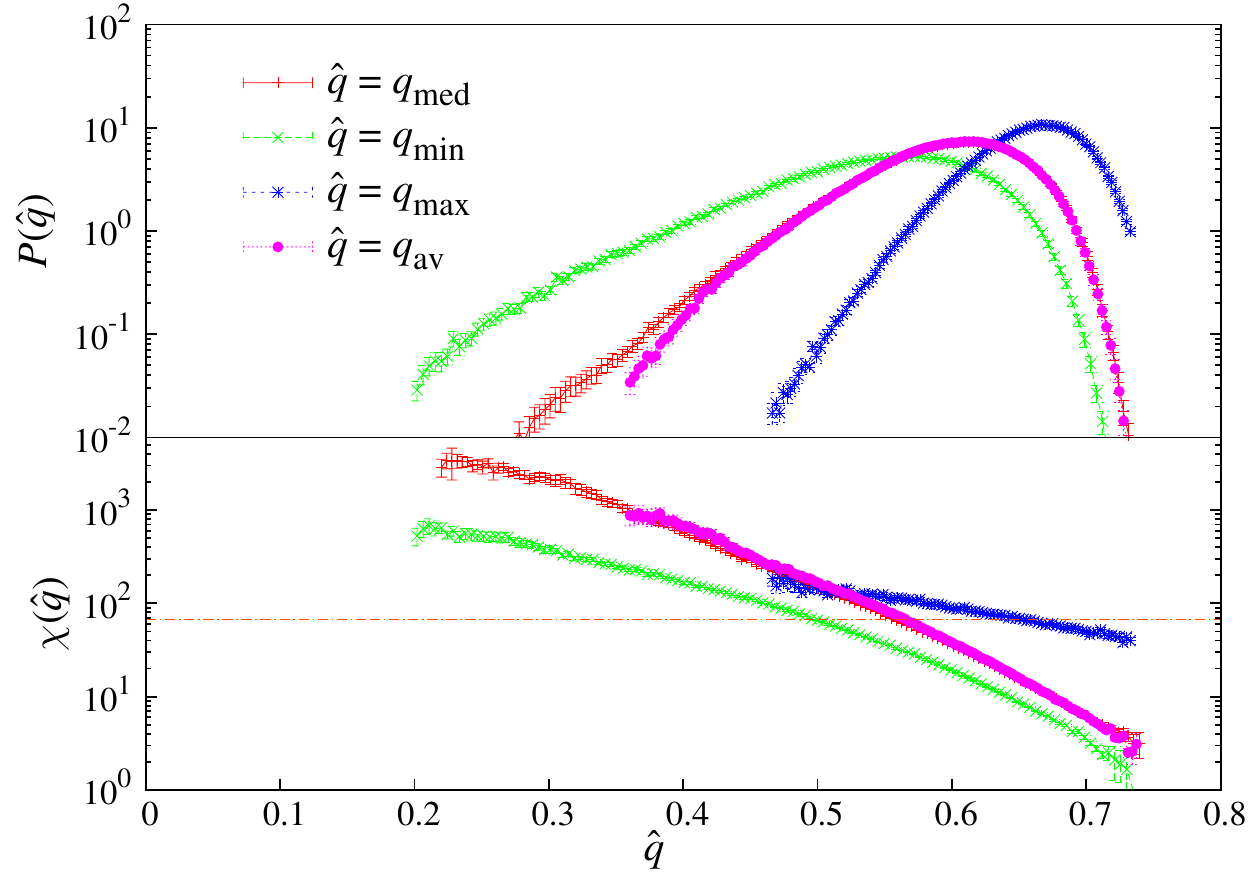}
\caption{Features of the diverse conditioning variates we proposed for
  $L=32$, $h=0.2$ and $T=0.805128$.  The top figure shows the
  histograms $P(\hat q)$ for our four candidates of conditioning
  variate: the minimum overlap $q_\mathrm{min}$ [of the six we can
    make with four replicas, recall (\ref{eq:cv})], the maximum
  $q_\mathrm{max}$, the median $q_\mathrm{med}$ and the average
  $q_\mathrm{av}$.  The histograms were constructed as explained in~\ref{app:hack}.
    The bottom figure depicts the size of the
  susceptibility $\chi$ for each value of the conditioning variate.
  The horizontal line marks the value of $\chi$ when it is averaged
  over the full set of measurements.  For aesthetic reasons in both
  figures we have cut the curves at the two end points, where they
  become extremely noisy due to poor sampling.  }
\label{fig:cv}
\end{figure}

For a qualitative description of the difference between the diverse
conditioning variates, in figure~\ref{fig:cv} (top) the reader can appreciate the
probability distribution functions for each of the
conditioning variates, while in figure~\ref{fig:cv} (bottom) we plotted the conditional
susceptibilities. 
From (\ref{eq:E-conditioned}) we stress that the integral of the values on
the top times the values of the bottom set yields the average susceptibility,
which is indicated with a horizontal line on the bottom plot of figure \ref{fig:cv}.
As it is also reflected by table \ref{tab:reglas-de-suma},
$q_\mathrm{max}$ is the worst conditioning variate, as its
$\chi$ does not vary much with the fluctuations of $q_\mathrm{max}$.
The steepest slope
is obtained when the conditioning variate is $q_\mathrm{av}$ or
$q_\mathrm{med}$, but the latter is smoother and covers a wider range of
$\chi$.

Figure~\ref{fig:cv} also displays the large deviations present in the system.
In fact one can see that the value of $q_\mathrm{med}$ at which the $P(q_\mathrm{med})$
has its maximum is significantly different with respect to the value
of $q_\mathrm{med}$ at which $\chi(q_\mathrm{med})$ assumes the value of the average.

\section{Quantiles and a modified finite-size scaling ansatz}
\label{sec:median}

We stated in section~\ref{sec:extended-abstract} that the set of
measurements with low $\hat q$ has a very different behaviour from the
measurements with high $\hat q$ (recall figure\ref{fig:Cr}). From now
on, we shall restrict ourselves to $\hat{q} = q_\mathrm{med}$, since
we evinced that the median is our best conditioning variate.  Our next
goal will be to carry out a finite-size scaling analysis based on the
$P(q_\mathrm{med})$ that lets us observe different parts of the
spectrum of behaviours of the system.

In order to analyse separately these different sets of measures, we
divide the $P(q_\mathrm{med})$ in 10 sectors, each containing $10\%$
of the measured $q_\mathrm{med}$.  We focus our analysis on the values
of $q_\mathrm{med}$ that separate each of these sectors. They are
called quantiles (see, e.g., \cite{hyndman:96}), and we label them
with the subscript $i=1,\ldots,9$. If we call $\tilde{q}_i(h,T,L)$ the
value of the $i^\mathrm{th}$ quantile, we can define it in the
following implicit way:
\begin{equation} 
\int_{-1}^{\tilde q_i} \mathrm{d}\hat q P(\hat q) = \frac{i}{10}\,.  
\end{equation} 
In~\ref{app:hack} we explain how $\tilde{q}_i(h,T,L)$ was computed.

We can adapt to the $i^\mathrm{th}$ quantile the definitions we gave 
in section~\ref{sec:obs}:
\begin{eqnarray}
  \chi_{\mathrm{R},i}(\mathitbf{k}) &=& \frac{1}{V} E\bigg(|\hat {\Phi}_{\mathitbf{k}}^{(ab;cd)}|^2 ~\bigg|~ \tilde q_i\bigg)\,,\\[2ex]
 \xi_{L,i}                          &=& \frac{1}{2 \sin{(k_\mathrm{min}/2)}} \sqrt{\frac{\chi_{\mathrm{R},i}(\bf{0})}{\chi_{\mathrm{R},i}(2\pi/L,0,0)}-1}\,,\\[2ex]
 R_{{12},i}                         &=& \frac{\chi_{\mathrm{R},i}(2\pi/L,0,0)}{\chi_{\mathrm{R},i}(2\pi/L,\pm2\pi/L,0)}\,.
\label{eq:obs_sep}
\end{eqnarray}
This way we can extend the finite-size scaling methodology to the $i^\mathrm{th}$ quantile:
\begin{equation}
 \left.\frac{\xi_L}{L}\right|_{T,h,L,i} = {\cal F}_{\xi_i}\left(L^{1/\nu}(T-T_\mathrm{c})\right)~~,~~\left.R_{12}\right|_{T,h,L,i} = {\cal F}_{R_i}\left(L^{1/\nu}(T-T_\mathrm{c})\right)\,.
\end{equation}
This is a new approach for finite-size scaling. 
Although it demands a very large amount of data because it is
done over a small fraction of the measurements, it allows us to perform finite-size scaling
on selected sets of measurements. Let us stress that no \emph{a priori} knowledge is required on
the probability distribution function $P(q_\mathrm{med})$: Quantiles are 
conceived in order to define a scaling that self-adapts when the volume increases.
In~\ref{app:h0} we show the validity of this ansatz in zero field.

\subsection{The \texorpdfstring{$P(q_\mathrm{med})$}{P(qmed)}}

Up to our knowledge, despite its simplicity the median overlap
$q_\mathrm{med}$ has not been studied before. In fact, we just lacked
the motivation to investigate its features. Yet, now we base our
analysis on this quantity, so we feel that it is necessary to dedicate
it a paragraph.

By its definition, the probability distribution $P(q_\mathrm{med})$
of the median overlap has narrower tails than $P(q)$ (recall
figure~\ref{fig:Pq}), although from figure~\ref{fig:cv} (top) it is clear that the
strong fluctuations persist also with $q_\mathrm{med}$.

\begin{figure}[t]
\includegraphics[width=\columnwidth]{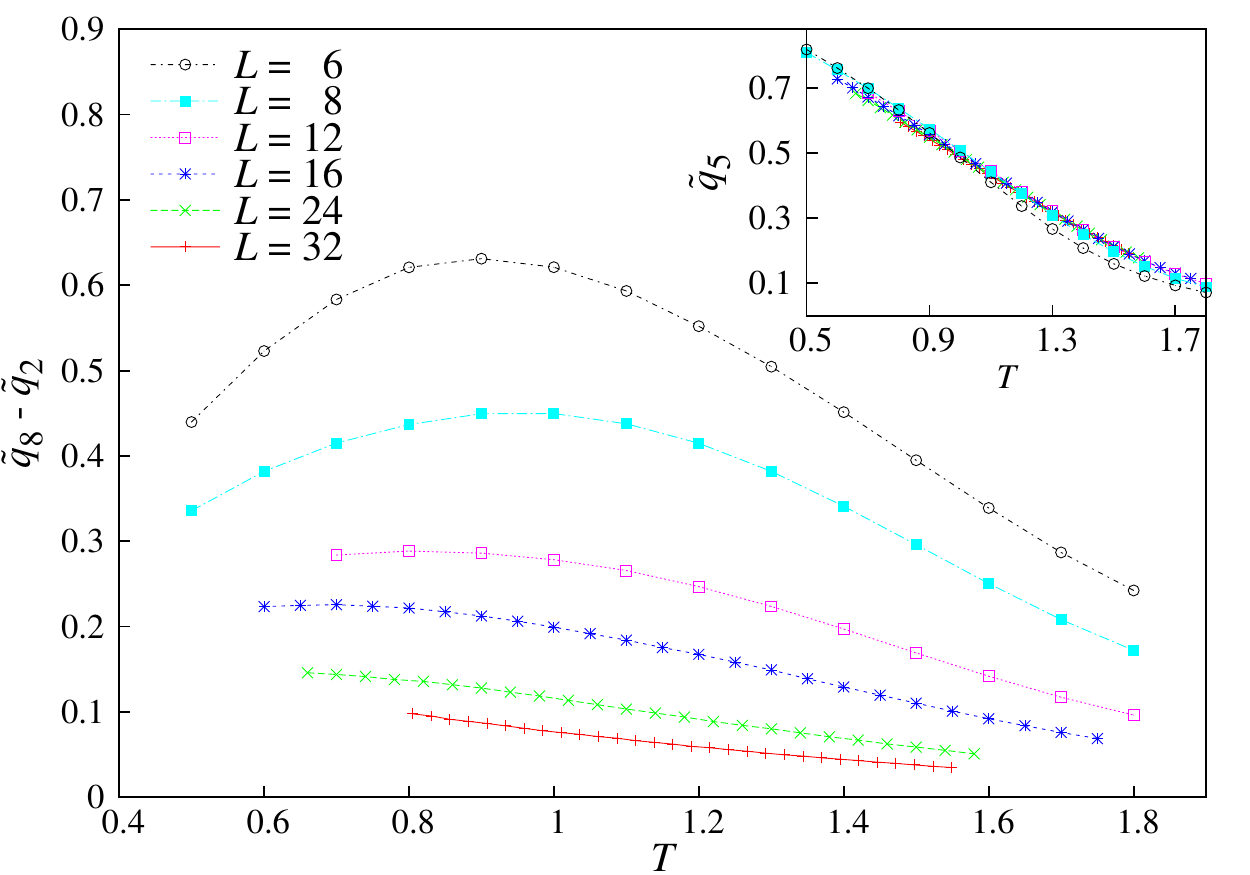}
\caption{Using $q_\mathrm{med}$ as conditioning variate, we show the
temperature dependence of the difference between quantiles $\tilde q_8-\tilde
q_2$, for all our lattice sizes, in a field of intensity $h=0.2$.  This
corresponds to the width of the central $60\%$ of area of $P(q_\mathrm{med})$.
This quantity can reveal a phase transition, since in the paramagnetic phase
the $P(q_\mathrm{med})$ should be a delta function, while in the spin-glass
phase it should have a finite support. We show the central $60\%$ and not a
wider range because it is an equivalent indicator of the phase transition,
and it is safer from rare events that would vanish in the
thermodynamic limit.
In the inset we show the position of $5^{th}$ quantile as a function
of temperature in all our lattice sizes. It is a very smooth curve with 
very small finite-size effects.}
\label{fig:anchura}
\end{figure}

The median of $P(q_\mathrm{med})$ corresponds to the fifth quantile. We will
prefer to call it ``$5^\mathrm{th}$ quantile'' rather than ``median of the
median overlap''.  Of the nine studied quantiles it is the smoothest and has
the least finite-size effects, as one can see from figure~\ref{fig:anchura}
(inset). Further analysis is given in~\ref{app:q_L}.

We remark also that the separation between the different $\tilde{q}_i$'s can be
used as order parameter, since its thermodynamic limit should be zero in the
paramagnetic phase, and greater than zero in the possible low-temperature phase
due to the (would-be) replica symmetry breaking.  Figure~\ref{fig:anchura} shows
the difference between the $8^\mathrm{th}$ and the $2^\mathrm{nd}$ quantile,
i.e., the $q_\mathrm{med}$-span of the central $60\%$ of the data. If we were
able to extrapolate a clean $L\rightarrow\infty$ limit for this curve, we would
be able to answer to whether the transition exists or not.  Unfortunately, even
for $T>T_\mathrm{c}(h=0)=1.1019(29)$, where we know that we are in the
paramagnetic phase, it is not possible to make good extrapolations since the
trend is strongly non-linear.  In~\ref{app:q_L} we show that extrapolations to
the thermodynamic limit were only possible in the trivial case of $h=0.4$ (deep
paramagnetic phase), and that between all the quantiles, the median curve is
the one that shows less finite-size effects.

\section{Results}
\label{sec:results}
\begin{figure}[t]
\centering
 \includegraphics[width=\columnwidth]{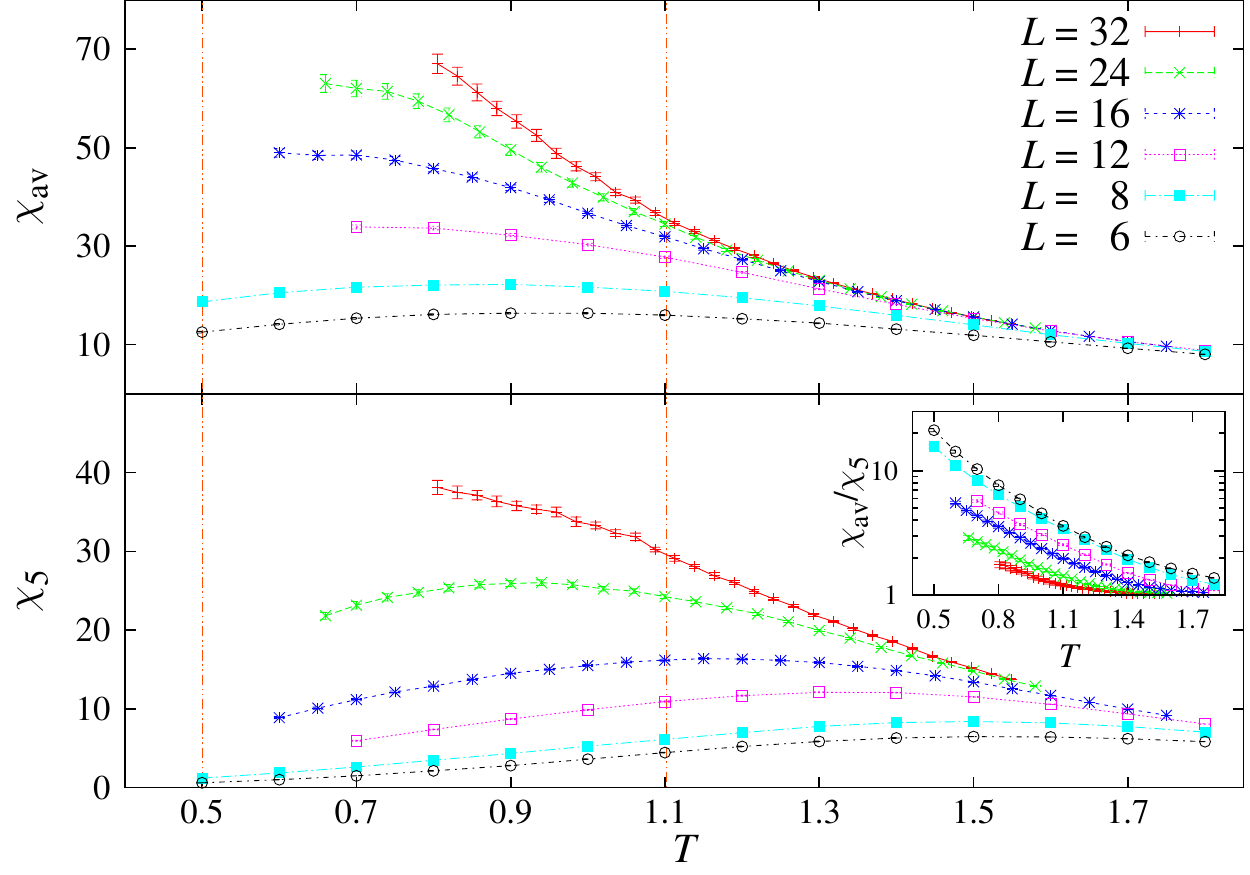}
\caption{The replicon susceptibility $\chi$ as a function of the
  temperature, for all the simulated lattice sizes and the field
  $h=0.2$.  We represent its average $\chi$ (top), and the
  $5^\mathrm{th}$ quantile $\chi_5$ with $\hat q=q_\mathrm{med}$
  (bottom). In both plots, the two vertical lines represent the upper
  bound of the possible phase transition $T^\mathrm{up}(h=0.2)=0.5$
  \cite{janus:14b}, and the zero-field critical temperature
  $T_\mathrm{c}(h=0)=1.109(29)$ \cite{janus:13b}.  The amplitudes and
  the positions of the peaks of $\chi$ are strikingly different (mind
  the different scales in the $y$ axes). The inset shows the ratio
  between the two, which we expect to tend to an order one constant in
  the thermodynamic limit. This is actually what we see at high
  temperatures.}
\label{fig:chi} 
\end{figure}

As already stressed, the behaviour of the system is characterised by very strong
fluctuations, and a wide and asymmetric $P(q)$. As a result, the average and
median behaviour are very different. 
In figure~\ref{fig:chi}, we show the replicon susceptibility: its average $\chi$ on the left
plot, and its fifth quantile $\chi_5$. Motivated by the arguments in section~\ref{sec:conditional}
all the quantiles we show in this section use the conditioning variate $\hat q=q_\mathrm{med}$.

Visibly, not only is the average susceptibility much larger than the
$5^\mathrm{th}$ quantile, but also the two have peaks at different
temperatures. Also, finite-size effects are much stronger in the case
of $\chi_5$ (yet, recall the inset in figure~\ref{fig:anchura},
finite-size effects on $\tilde q_5$ are tiny).\footnote{ We made power 
law extrapolations to $L\rightarrow\infty$ of the maxima of the susceptibility, but they 
were not satisfactory (too large $\chi^2/\mathrm{DOF}$). Only for $h=0.2,0.4$ were we able to fit the maxima's
heights and obtained $\eta(h=0.2) \approx 0.6$ and $\eta(h=0.4)\approx 0.9$.}

\begin{figure}[tbh]
\centering
\includegraphics[width=\columnwidth]{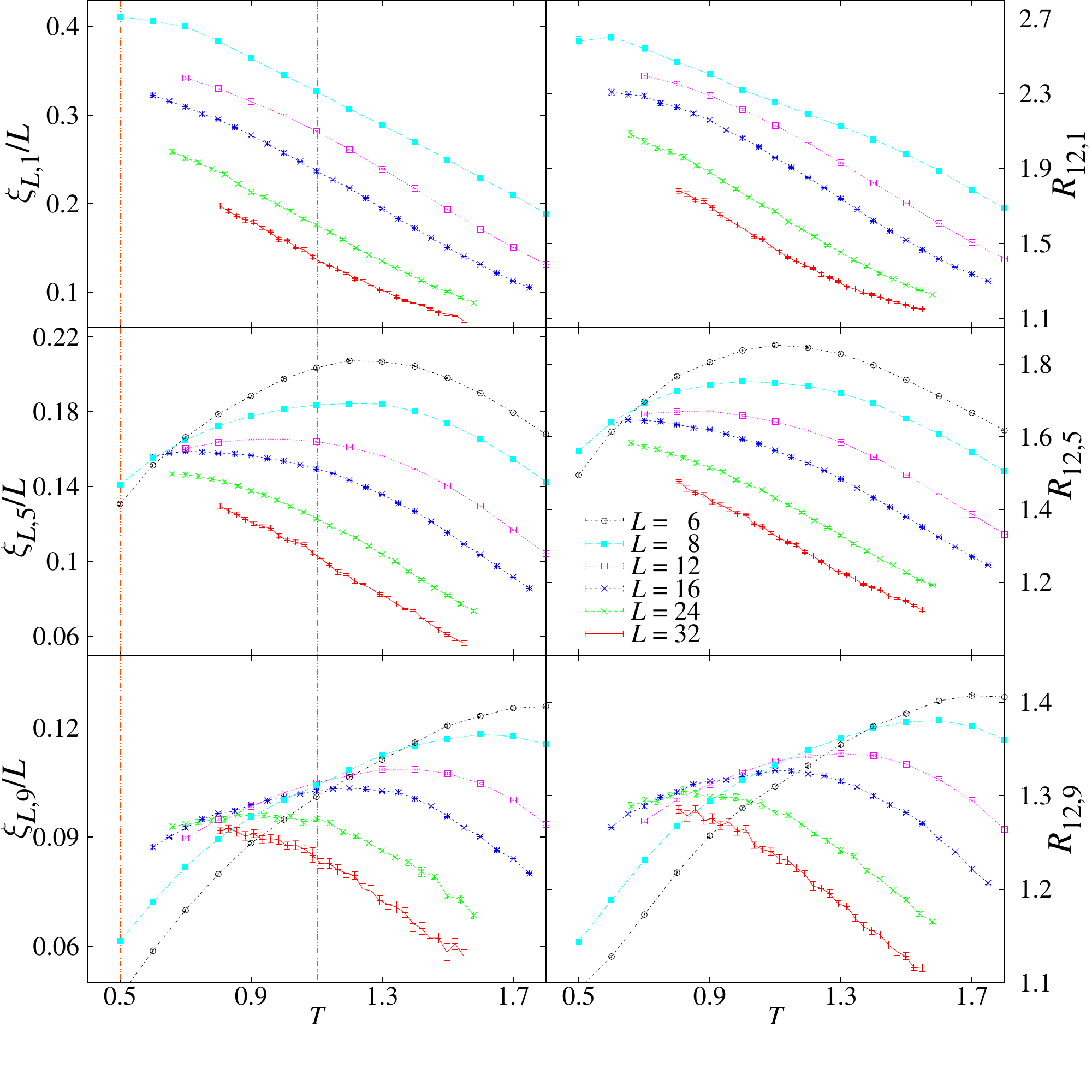}
\caption{Finite-size indicators of a phase transition, computed for
  $h=0.2$. On the left side we plot, for quantiles 1 (top), 5 (middle)
  and 9 (bottom), the correlation length in units of the lattice size
  $\xi_L/L$ (left) versus the temperature, for all our lattice sizes
  except $L=6$ (we show in~\ref{app:Pq} that the quantile description
  is not suitable for $L=6$ because there is a double peak in the
  $P(q)$).  On the right we show analogous plots, for $R_{12}$.  The
  vertical line on the left marks the upper bound $T^\mathrm{up}$ for
  a possible phase transition given in \cite{janus:14b}, while
  the one on the right marks the zero-field transition temperature
  $T_\mathrm{c}$ given in \cite{janus:13b}.  Quantile 1 has the
  same qualitative behaviour of the average $\xi_L/L$, shown in
  figure~\ref{fig:xiL}, while quantiles 5 and 9 suggest a scale
  invariance at some temperature $T_h<T^\mathrm{up}$.}
\label{fig:xiL-separil_h02}
\end{figure}

We show in figure~\ref{fig:xiL-separil_h02} how sorting the data with the quantiles
revealed the presence of different types of behaviour, by plotting the
$\xi_L/L$ and the $R_{12}$ for quantiles 1, 5 and 9 at $h=0.2$.\footnote{
In \ref{app:h01} we show the same plots for $h=0.1$, and in \ref{app:h0} 
we validate our method in the null-field case, where it is
known that there actually is a phase transition.} There are 
two vertical lines in each figure. The one
on the left represents the upper bound $T^\mathrm{up}(h)$ for the phase
transition (meaning that no phase transition can occur for $T>T^\mathrm{up}(h)$) 
given in \cite{janus:14b}, while the one on the right
indicates the zero field critical temperature
$T_\mathrm{c}=1.1019(29)$ \cite{janus:13b}. 

We can see that the $1^{st}$ quantile
has the same qualitative behaviour of the average (figure \ref{fig:xiL}), but
lower values, since the main contribution to the average comes from data whose
$q_\mathrm{med}$ is even lower than $\tilde q_1$.
Moreover, one can notice that in figure~\ref{fig:xiL} the indicators $\xi_L/L$ and
$R_{12}$ show a different qualitative behaviour when the lattices are small ($R_{12}$
shows a crossing). This discrepancy vanishes when we look only at the first quantile:
Separating different behaviours enhances the consistency between $\xi_L/L$ and
$R_{12}$.

The behaviour of the $5^\mathrm{th}$ quantile
is quite different, since now it appears reasonable that the curves cross at
some $T\lesssim T^\mathrm{up}(h)$. The crossings become even more evident when we
consider the highest quantile.

All this is consistent with the
arguments of section~\ref{sec:extended-abstract}, where we showed how the
correlation function is dominated by a little portion of data, near the first
quantile (figure~\ref{fig:Cr}), while the behaviour of the majority of the 
samples is hidden.

Unfortunately, the high non-linearity of the curves impedes an
extrapolation of the crossing points, but they are apparently compatible with
the upper bound $T^\mathrm{up}$, and their heights apparently do not depend
on the intensity of the applied field $h$ (\ref{app:h}).

\section{This is not an echo of the \texorpdfstring{\boldmath $h=0$}{h=0} transition}
\label{sec:zero-field}
The crossing suggested by the quantiles 5 and 9 in figure~\ref{fig:xiL-separil_h02} is unlikely
to be caused by the zero-field transition, since it appears at
$T<T_\mathrm{c}$, and shifts towards lower temperatures as the lattice size
increases. Also, the value of $\xi_L/L$ ($R_{12}$) at the possible crossing point of the fifth quantile
is upper-bounded to $\xi_L/L\simeq0.16$ ($R_{12}\simeq1.65$), while for $h=0$ it is considerably larger 
($\xi_L(T_\mathrm{c})/L \simeq 0.28$ [$R_{12}(T_\mathrm{c})\simeq2.15$]), see ~\ref{app:h0}.
In this section we will give more arguments sustaining that
what is seen is not an effect of the zero-field transition.

\subsection{An escaping transition}
As pointed out in section~\ref{sec:extended-abstract}, there is a controversy
because we observe a wide $P(q)$, just like in the mean-field
model, but the curves $\xi_L/L(T)$ and $R_{12}(T)$ do not show any sign of a
crossing. If we were in the presence of a phase transition, a straightforward
explanation could reside in an anomalous exponent $\eta$ close to 2 \cite{baityjesi:14}, since at
the critical temperature the replicon susceptibility scales as $\chi_\mathrm{R}(L)\sim
L^{2-\eta}$ (\ref{eq:chi-scaling}). 
It is possible to calculate $\eta$ with the quotients method \cite{nightingale:76, ballesteros:96}, 
by comparing the susceptibility of different lattice sizes at the critical point $T^{*}$:
\begin{equation}
 \label{eq:quotients}
 \frac{\chi_{2L}(T^*)}{\chi_L(T^*)} = 2^{2-\eta} + \ldots\,,
\end{equation}
where the dots stand for subleading terms.
This definition only makes sense at criticality, but we can extend 
it in an effective manner to a generic temperature. This way we can delineate 
an effective exponent
\begin{equation}
\label{eq:eta-eff}
 \eta_\mathrm{eff}(T;L,2L) = 2 - \log_2 \frac{\chi_{2L}(T)}{\chi_L(T)} \,.
\end{equation}
In case there were a phase transition at a finite temperature $T_{h}$,
we would have $\eta_\mathrm{eff}(T_{h})=\eta$. We should have
$\eta_\mathrm{eff} = 2$ in the paramagnetic phase, $\eta_\mathrm{eff}
= -1$ in the deep spin-glass phase [see~\ref{app:eta-eff}, keeping in
mind that $\eta_\mathrm{eff} = -1$ is somewhat trivial in the limit
$h\to 0$, where $\chi$ reduces to $\chi= V E(q^2)$] and signs of a
crossing at $\eta_\mathrm{eff} =\eta(h=0) = -0.3900(36)$
\cite{janus:13b} in the limit of a complete domination by the $h=0$
transition.

\begin{figure}[tbh]
\centering
  \includegraphics[width=0.75\columnwidth]{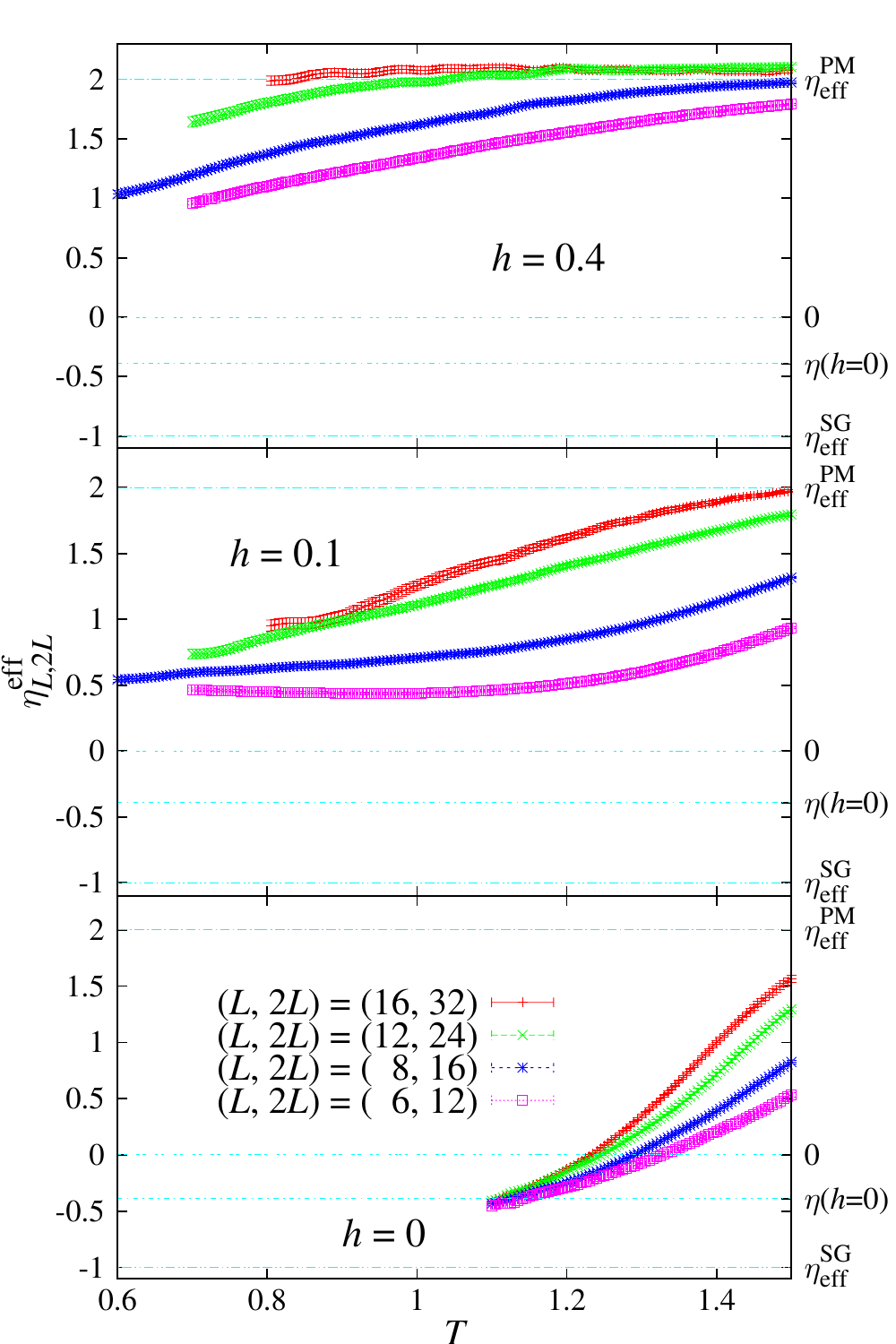}
\caption{We plot $\eta_\mathrm{eff}(T)$, defined in
(\ref{eq:eta-eff}), for all the pairs $(L,2L)$ we could form. 
The magnetic fields are $h=0.4$ (top), $h=0.1$ (centre) and $h=0$ (bottom).
The $h=0$ data comes from \cite{janus:13b}.
In each plot we uses horizontal lines to underline meaningful limits, and we
label them with a tic on the right axis.
From up to down, we depict the limit
$\eta_\mathrm{eff}^\mathrm{PM}=2$ of a system in the paramagnetic phase, the $\eta_\mathrm{eff}=0$ axis,
the zero-field value
$\eta_\mathrm{eff}(h=0,T_\mathrm{c})=-0.3900(36)$ \cite{janus:13b},
and its value in a deep spin-glass phase $\eta_\mathrm{eff}^\mathrm{SG}=-1$.
Notice the difference between the case with or without a field.
For $h=0.1$ the curves appear to converge to
a positive $\eta_\mathrm{eff}\simeq0.5$, while in the latter all the curves become negative
and merge at $\eta_\mathrm{eff}(h=0,T_\mathrm{c})$.} 
\label{fig:eta-eff}
\end{figure}

In figure~\ref{fig:eta-eff} we show $\eta_\mathrm{eff}(T)$ for $h=0.4,
h=0.1$, and $h=0$ (the $h=0$ data come from the simulations we
performed in \cite{janus:13b}).\footnote{ For each jackknife block we
  calculated $\eta_\mathrm{eff}(T)$ and made a cubic spline
  temperature interpolation.}
If a phase transition were present, but hidden by heavy finite-size effects, 
we would expect at least that the
$L$-trend of $\eta_\mathrm{eff}$ be decreasing. Contrarily, the larger our
lattices, the wider the temperature range in which $\eta_\mathrm{eff} = 2$.
The apparent phase transition shifts towards lower temperature when we
suppress finite-size effects. The data in our possession is not enough to state
whether this shift will converge to a positive temperature.  In any case, this is
compatible with the upper bounds to a possible transition given in \cite{janus:14b}.

On the other side, $\eta_\mathrm{eff}$ stays positive for all
our simulated lattices (except $h=0.05$, $L=6$), and  that even for
$T<T_\mathrm{c}(h=0)$ it tends to some value around $0.5$, so it is unlikely
that the null field transition is dominating the system's behaviour.

\subsection{Scaling at \texorpdfstring{$T=T_\mathrm{c}(h=0)$}{T=Tc(h=0)}}
\begin{figure}[tbh]
 \includegraphics[width=\columnwidth]{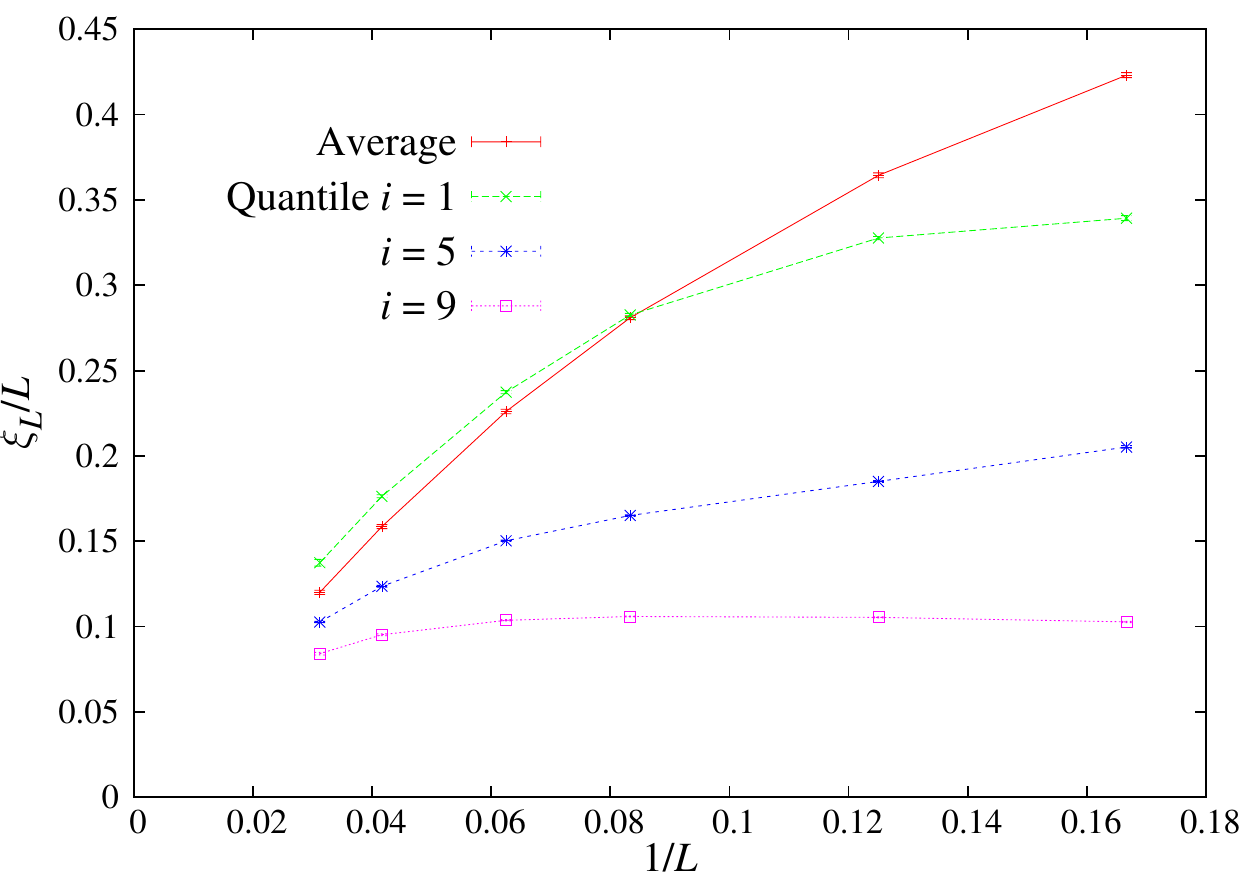}
 \caption{Scaling of $\xi_L/L$ at the null-field critical temperature
$T_\mathrm{c}=1.109(29)$ \cite{janus:13b}, with $h=0.2$. We show the behaviour of the average,
and of quantiles 1, 5 and 9.  If $L$ is large enough, $\xi_L/L$ should go as
$1/L$, while if the system is seeing purely an echo of the divergence of the $h=0$ transition
transition, then $\xi_L/L$ should be constant.}
\label{fig:xiL_Tc}
\end{figure}

From the scaling with the lattice linear size of $\xi_L/L$ at
$T_\mathrm{c}=T_\mathrm{c}(h=0)$, we can get another element to discard the
hypothesis that the $h=0$ transition is biasing significantly our measures.
Assuming that there is no critical line for $h>0$, a very large correlation
length could be due to an echo of the zero-field transition or a low-temperature 
effect. 
In a theory that predicts that system is critical only at $h=0$, $T=T_\mathrm{c}$,
the effects of this echo on the $h>0$ behaviour should be maximal near $T=T_\mathrm{c}$.
So, if we find a $\xi$ that is large compared to our lattice sizes
for $T<T_\mathrm{c}$, a primary check is to monitor the scaling of the coherence
length at $T_\mathrm{c}$.
figure~\ref{fig:xiL_Tc} shows the scaling of $\xi_L/L$ at
$T_\mathrm{c}$ with $h=0.2$. We plot the average, the first, the fifth and the
highest quantile. All of them show a clear decrease of $\xi_L/L$ when
increasing the lattice size, so our lattice sizes are large enough to
state that the divergence at $h=0$ is not dominating $\xi_L$'s behaviour.
On the other side, we are still far from the thermodynamic limit, 
since when the lattices
are large enough, $\xi_L(T_\mathrm{c})/L$ should decay to zero linearly in
$1/L$. 

\section{Conclusions}
\label{sec:conclusions}

We have studied the equilibrium behaviour of the three-dimensional Ising
Edwards-Anderson spin glass in an external magnetic field.  Thermalising the
system at sufficiently low temperature was a computationally hard task and
required the use of the \emph{Janus} dedicated computer to thermalise lattice
sizes up to $L=32$, down to temperatures $T\geq0.8$. 

First of all, we carried out a traditional analysis of our data. We chose
observables that would be scale invariant at the critical temperature, and
compared them for different lattice sizes, looking for crossings in their
temperature curves.  With this procedure we found no traces of a phase
transition. 

Yet, the scenario is more complicated. Despite the absence of crossings, indications
that something non-trivial is going on are given by signals such as a growing
correlation length (even for our largest lattices), peaks in the susceptibility,
and a wide probability distribution function of the overlap.

We noticed a wide variety of behaviours
within the same set of simulation parameters. Some measurements presented
signs of criticality, while others did not. So, we tried to classify them in a
meaningful way.  We sorted our observables with the help of a
conditioning variate, and came up with a quantitative criterion to select the
best conditioning variate.  Between the ones we proposed, the function of the
instant overlaps that made the best conditioning variate turned out to be the
median overlap $q_\mathrm{med}$.

As a function of the median overlap, the scenario appeared rather
non-trivial.  The averages turned out to be dominated by a very small
number of measurements.  Those with a small $q_\mathrm{med}$ behaved
similarly to the average: long correlation lengths, very large
susceptibilities, and no signs of criticality.  On the other side, the
median behaviour was far from the average, and the behaviour of most of
the measurements was qualitatively different from the average, with
smaller correlation lengths and susceptibilities, but non-negligible
indications of scale invariance right below the upper bound
$T^\mathrm{up}(h)$ given in \cite{janus:14b}. Furthermore,
separating the different behaviours of the system we obtain mutually
consistent indications of criticality from our primary dimensionless
magnitudes $\xi_{L,i}/L$ and $R_{12,i}$. The achievement of this
consistency is an important step forward with respect to
\cite{janus:12}, where the phase transition was revealed only
by the $R_{12}$ indicator, but it was invisible to $\xi_L/L$.

Unfortunately we were not able to make a quantitative prediction on
the critical temperatures $T_\mathrm{c}(h)$, because the observables
as a function of the lattice size and of the temperature were very
nonlinear, and the temperatures we reached were not low enough reliably to
identify the crossing points of the quantile-dependent
$\xi_{L,i}/L$ and $R_{12,i}$.

Overall, the presence of a phase transition appears plausible from our
simulations (see also \ref{app:HT}).  Perhaps more importantly, now
the challenge is well defined: in order to be able to give,
numerically, a conclusive answer on the presence of a de
Almeida-Thouless line we need push our simulations down to
$T\simeq0.4$ (at $h=0.2$).  We believe that \emph{Janus II}, the next
generation of our dedicated computer \cite{janus:14b}, will be able to
assume this challenge.

\section*{Acknowledgments}

We thank M.A. Moore and J. Yeo for pointing out to us the potential
usefulness of the Fisher-Sompolinsky scaling.

The Janus project has been partially supported by the EU (FEDER funds,
No.  UNZA05-33-003, MEC-DGA, Spain); by the European Research Council
under the European Union's Seventh Framework Programme (FP7/2007-2013,
ERC grant agreement no.  247328); by MINECO (Spain) (contracts
FIS2012-35719-C02, FIS2010-16587, TEC2010-19207); by the SUMA project
of INFN (Italy); by the Junta de Extremadura (GR10158); by the
European Union (PIRSES-GA-2011-295302).  F.R.-T. was supported by the
Italian Research Ministry through the FIRB project No. RBFR086NN1;
M.B.-J. was supported by the FPU program (Ministerio de Educacion,
Spain); R.A.B. and J.M.-G. were supported by the FPI program
(Diputacion de Aragon, Spain); S.P.-G. was supported by the ARAID
foundation; finally J.M.G.-N. was supported by the FPI program
(Ministerio de Economia, Spain).

\appendix

\newcommand{\nocontentsline}[3]{}
\newcommand{\tocless}[2]{\bgroup\let\addcontentsline=\nocontentsline#1{#2}\egroup}

\tocless\section{Quantiles for different fields \label{app:h}}
\addcontentsline{toc}{section}{Appendix A. Quantiles for different fields}

\tocless\subsection{Quantiles for \texorpdfstring{$h=0.1$}{h=0.1}\label{app:h01}}
\begin{figure}[t]
\centering
 \includegraphics[width=\columnwidth]{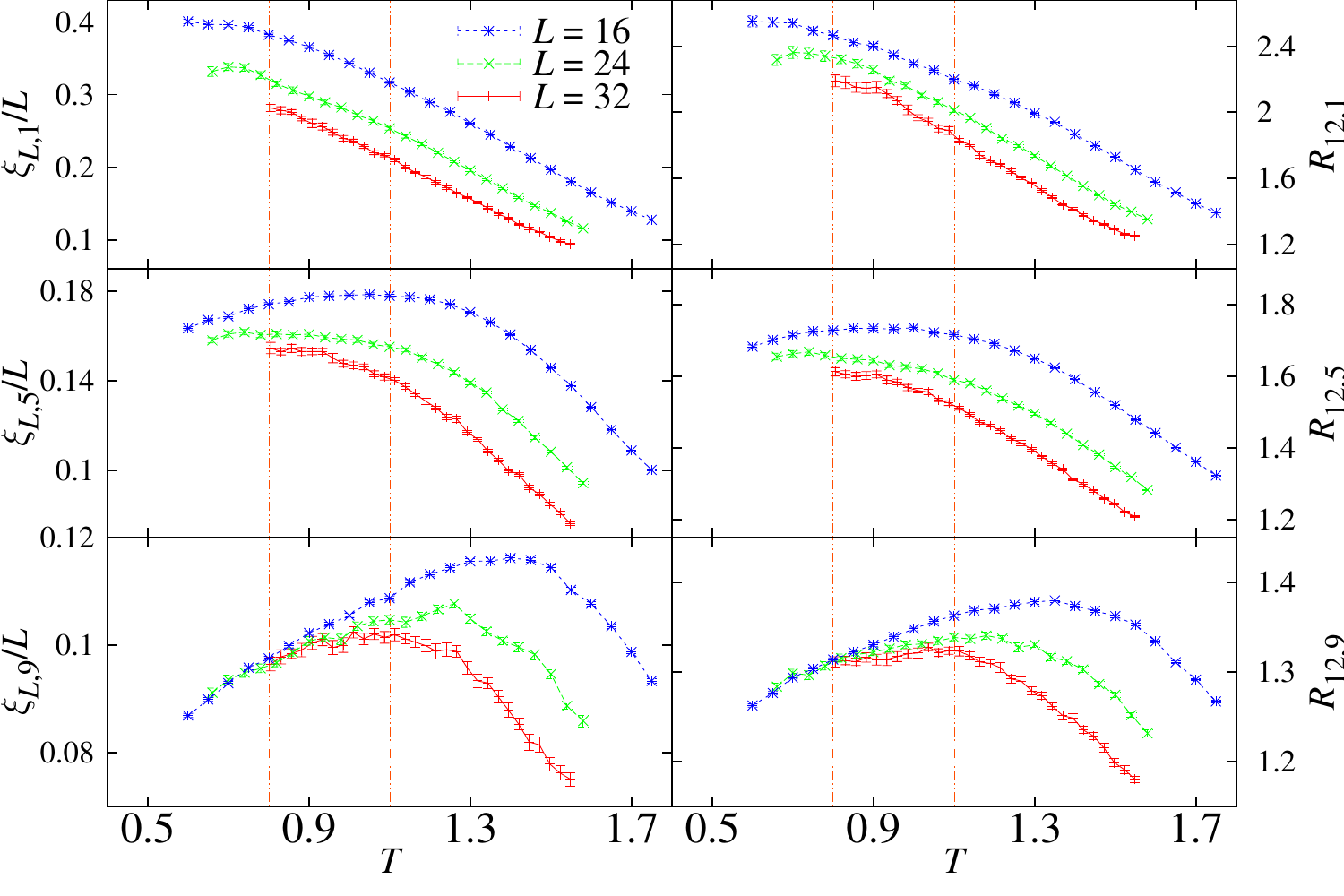}
\caption{Same as figure~\ref{fig:xiL-separil_h02}, but for $h=0.1$. This time
the effects of the zero-temperature transition are stronger, so we removed from
the plot sizes $L=6,8,12$. In~\ref{app:Pq} we show
that the quantile description is not suitable for smaller
lattices due to crossover effects from the zero-field behavior.}
\label{fig:xiL-separil_h01}
\end{figure}

The careful reader might have noticed that the upper bound $T^\mathrm{up}(h)$
for the possible phase transition given in \cite{janus:14b} is higher when
the field is lower: $T^\mathrm{up}(0.1) = 0.8 > T^\mathrm{up}(0.2)$. It is then
justified to ask oneself how do the quantile-plots look like for $h=0.1$.  We
show them in figure~\ref{fig:xiL-separil_h01}.
Since the field is lower, the effects on the double peak on the first quantile
(\ref{app:Pq}) extend to larger lattices than for $h=0.2$. Thus, we
show only the non-biased sizes, i.e., $L>12$. 

Although the $9^\mathrm{th}$ quantile shows signs of scale invariance
at $T=T^\mathrm{up}(0.1)$, the behavior of the $5^\mathrm{th}$
quantile suggests a scale invariance around $T=0.5$.  We believe that
the $5^\mathrm{th}$ quantile is a better indicator, since the position
of the fifth quantile $\tilde q_5$ has less finite-size effects (it
practically has none, figure~\ref{fig:anchura}--inset) than $\tilde
q_9$.

It is interesting to focus on the height of the crossings of each
quantile, and compare them with $h=0.2$ (recall
figure~\ref{fig:xiL-separil_h02}). This is expected to be a universal
quantity, and in the hypothesis of a phase transition it should be the
same for both fields.  Although it is not possible to assign error
bars to the these values, it is possible to see that both for $h=0.1$
and $h=0.2$ the heights are similar
($\xi_{L,5}/L\approx0.15$, $\xi_{L,9}/L\approx0.09$,
$R_{12,5}\approx1.6$, $R_{12,9}\approx1.3$).\\

\tocless\subsection{Quantiles for \texorpdfstring{$h=0$}{h=0}\label{app:h0}}

\begin{figure}[t]
\centering
 \includegraphics[width=\columnwidth]{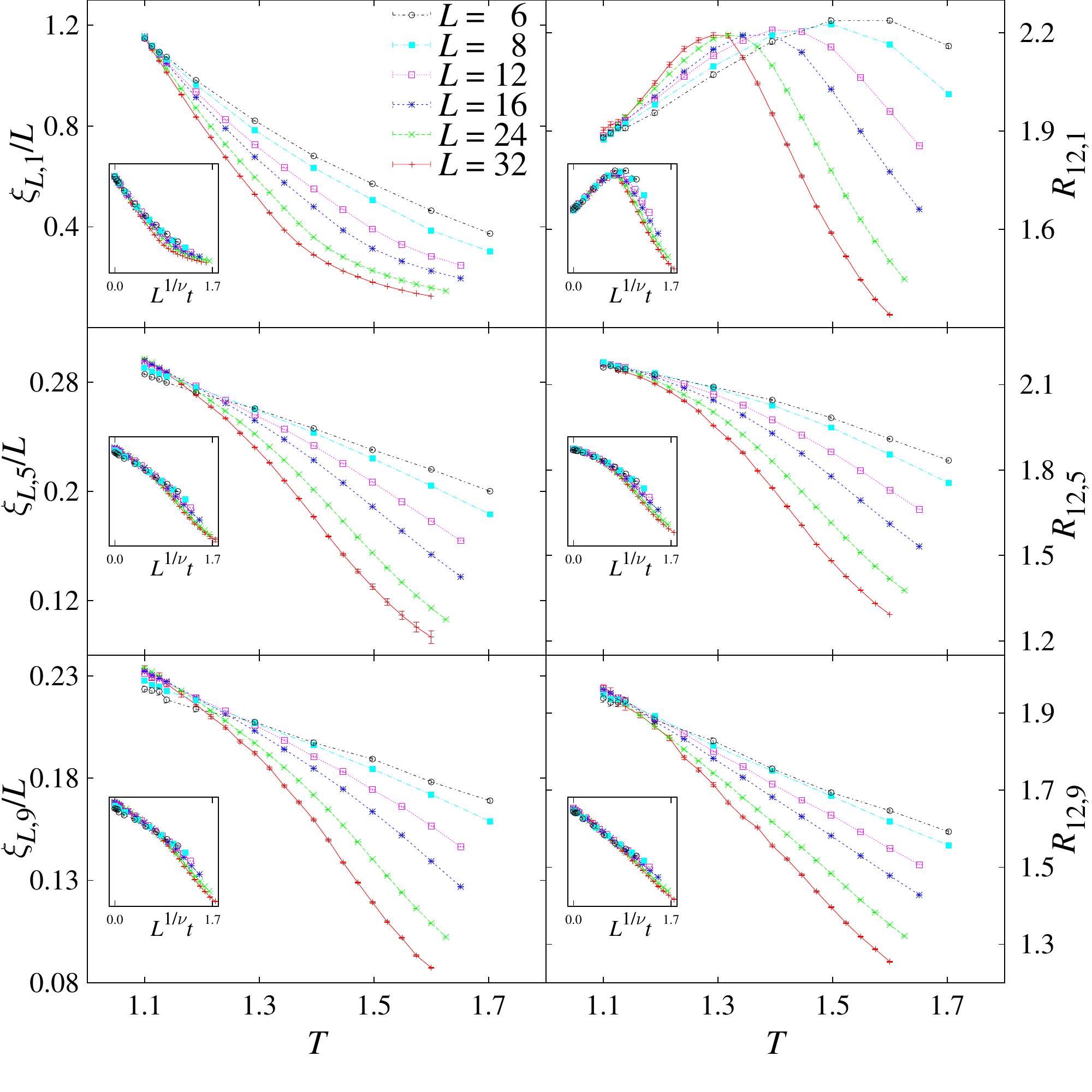}
 \caption{Same as figure~\ref{fig:xiL-separil_h02}, but for $h=0$. The data come from \cite{janus:13b}.
We used 256000 samples for each lattice size.
The insets show the same data of the larger sets, but as a function of the scaling variable $L^{1/\nu}t$,
where $t$ is the reduced temperature $t=(T-T_\mathrm{c})/T_\mathrm{c}$.}
 \label{fig:xiL-separil_h0}
\end{figure}

We take advantage of our $h=0$ data from \cite{janus:13b} to validate the
quantile description by showing its behavior in the zero-field case. 
Two replicas would be enough to construct connected correlators in $h=0$,
and using the 4-replica definitions proposed in section~\ref{sec:obs} only
adds noise to the results. Yet, we opted for the latter option
because the objective of the current section is the validation of the 
full procedure proposed herein.

In the absence of a magnetic field
we expect that the curves $\xi/L(T)$ and $R_{12}$ cross no matter the quantile, since
the behavior of the system is not dominated by extreme events and crossover fluctuations.
Also, in this case the data in our hands arrive down to the critical point, so the crossings
ought to be visible.

In fact, one can see from figure \ref{fig:xiL-separil_h0} that all the quantiles 
show visible signs of a crossing at $T_\mathrm{c}$ both in the case of $\xi_L/L$
and of $R_{12}$. 
Furthermore, if we plot the same data as a function of the scaling variable 
$L^{1/\nu}(T-T_\mathrm{c})/T_\mathrm{c}$ the data collapses well for all the quantiles 
(figure~\ref{fig:xiL-separil_h0}, insets).

\begin{figure}[bhtp]
\centering
\includegraphics[width=\columnwidth]{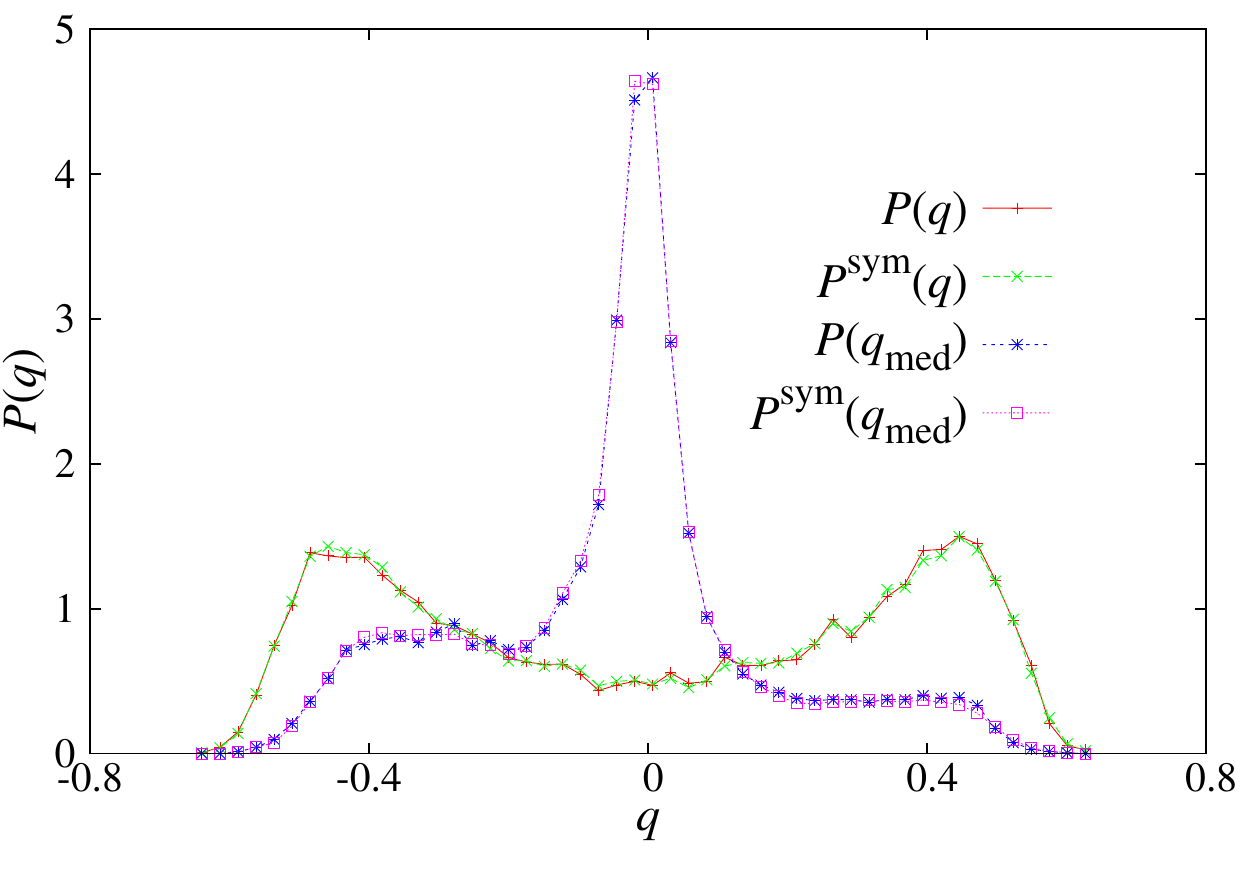}
 \caption{Probability distribution functions for $h=0$, $L=24$, $T=1.1$. The data come from 512 samples where we
 took all the $16^4$ combinations of overlaps per sample.
 We show $P(q)$, that in null field
 is symmetric, and $P(q_\mathrm{med})$, that is not. We also plot the symmetrised histograms $P^\mathrm{(sym)}(q)$
 and $P^\mathrm{(sym)}(q_\mathrm{med})$.}
 \label{fig:Pq_h0}
\end{figure}

Some reader may be surprised that quantiles 1 and 9 show different
behavior, being $P(q)$ symmetrical (figure~\ref{fig:Pq_h0}). The reason
is that, although $P(q)$ is symmetrical, $P(q_\mathrm{med})$ is not.
In fact, given six overlaps $q^{ab}, q^{ac}, q^{ad}, q^{bc}, q^{bd},
q^{cd}$ coming from four configurations
$\{\sigma^{(a)}\}$,$\{\sigma^{(b)}\}$,$\{\sigma^{(c)}\}$,$\{\sigma^{(d)}\}$,
each enjoying a $Z_2$ symmetry, the distribution of their median
privileges negative values.\footnote{Let us give a simple example. Take 4 $Z_2$-symmetric single-spin systems that can assume different values 
$s_1=\pm1, s_2=\pm2, s_3=\pm3, s_4=\pm4$.
We can construct 6 overlaps $q_{ij}(s_1, s_2, s_3, s_4)$.
If we explicitate the $Z_2$ symmetry, taking all the combinations of our random variables, the histogram
of $q$ will be symmetric with zero mean. Yet, if we take the histogram of the median overlap, it will be asymmetric
with mean $\langle q_\mathrm{med}\rangle = -3$. This can easily be checked by computing all the possible combinations
of the signs of the $s_i$ and computing the median in each case: $q_\mathrm{med}(+1, +2, +3, +4) = 5$, 
$q_\mathrm{med}(+1, +2, +3, -4) = -1$, $q_\mathrm{med}(+1, +2, -3, -4) = -3.5$, and so on.
} We show this in figure~\ref{fig:Pq_h0}, where we give both the $P(q)$
and the $P(q_\mathrm{med})$ for $h=0$, $L=32$, $T=1.1$. The first is
symmetrical and the second is not.  To convince the reader that the
starting configurations do enjoy $Z_2$ symmetry, we also construct the
symmetrised functions $P^\mathrm{(sym)}(q)$ and
$P^\mathrm{(sym)}(q_\mathrm{med})$.  These two functions are obtained
by explicitly imposing the $Z_2$ symmetry: for each measurement we
construct the $2^4$ overlaps with both $\{\sigma_i\}$ and
$\{-\sigma_i\}$.  It is visible from figure~\ref{fig:Pq_h0} that
$P^\mathrm{(sym)}(q_\mathrm{med})$ is asymmetric even though we
imposed by hand the $Z_2$ symmetry on the configurations.\\

\tocless\section{4-Replica Correlators\label{app:GL}}
\addcontentsline{toc}{section}{Appendix B. 4-replica correlators}
In the presence of a magnetic field it is not possible to construct connected correlation
functions with the use of only two replicas because in the paramagnetic phase $\overline{\langle q\rangle} = q_\mathrm{EA}>0$.

With 4 replicas we can construct 3 different correlators at distance $\mathitbf{r}$.
\begin{eqnarray}
G_1(\mathitbf{r})&=\frac{1}{V}\sum_{\mathitbf{x}}\,\overline{\langle
  \sigma_{\mathitbf{x}} \sigma_{\mathitbf{x+r}} \rangle^2} \nonumber\\
&= \frac{1}{V}\sum_{\mathitbf{x}}\,\overline{\langle
  \sigma_{\mathitbf{x}}^{(a)} \sigma_{\mathitbf{x+r}}^{(a)} \sigma_{\mathitbf{x}}^{(b)} \sigma_{\mathitbf{x+r}}^{(b)} \rangle}
\,,\\
G_2(\mathitbf{r})&=\frac{1}{V}\sum_{\mathitbf{x}}\,\overline{\langle \sigma_{\mathitbf{x}} \sigma_{\mathitbf{x+r}} \rangle\langle \sigma_{\mathitbf{x}}\rangle \langle 
\sigma_{\mathitbf{x+r}}\rangle} \nonumber\\
&= \frac{1}{V}\sum_{\mathitbf{x}}\,\overline{\langle
  \sigma_{\mathitbf{x}}^{(a)} \sigma_{\mathitbf{x+r}}^{(a)} \sigma_{\mathitbf{x}}^{(b)} \sigma_{\mathitbf{x+r}}^{(c)} \rangle}\,,\\
G_3(\mathitbf{r})&=\frac{1}{V}\sum_{\mathitbf{x}}\,\overline{\langle
  \sigma_{\mathitbf{x}}  \rangle^2\langle \sigma_{\mathitbf{x+r}}
  \rangle^2}\\
&= \frac{1}{V}\sum_{\mathitbf{x}}\,\overline{\langle
  \sigma_{\mathitbf{x}}^{(a)} \sigma_{\mathitbf{x+r}}^{(b)} \sigma_{\mathitbf{x}}^{(c)} \sigma_{\mathitbf{x+r}}^{(d)} \rangle}\,.\nonumber
\end{eqnarray}

None of those goes to zero for large $\mathitbf{r}$, but, in the
paramagnetic phase, for large $\mathitbf{r}$ they all tend to the same
value, $q_\mathrm{EA}$. So, to create connected correlators, we can
make two linearly independent combinations of them, and obtain the
basic connected propagators of the replicated field theory
\cite{dedominicis:98}~\footnote{In the effective field theory the
  longitudinal ($G_\mathrm{L}$) and anomalous ($G_\mathrm{A}$)
  propagators are degenerated.}
\begin{equation}
\begin{array}{rrrrrrr}
G_\mathrm{R}&=&G_1&-&2G_2&+&G_3\,,\\
G_\mathrm{L}&=&G_1&-&4G_2&+&3G_3\,.
\end{array}
\label{eq:GRLA0}
\end{equation}
It is easy to check that these relations imply
\begin{eqnarray}
\label{eq:gamma2}
G_\mathrm{R} (\mathitbf{r}) &=& \overline{\left[\langle\sigma_\mathitbf{x}\sigma_{\mathitbf{x}+\mathitbf{r}}\rangle
					     -\langle\sigma_\mathitbf{x}\rangle\langle\sigma_{\mathitbf{x}+\mathitbf{r}}\rangle\right]^2}\,,\\[1ex]
G_\mathrm{L} (\mathitbf{r}) &=& 2 G_\mathrm{R} (\mathitbf{r}) - \Gamma_2 (\mathitbf{r})~~,~~~~~~~~~
			      \Gamma_2 (\mathitbf{r}) =\overline{\left[\langle\sigma_\mathitbf{x}\sigma_{\mathitbf{x}+\mathitbf{r}}\rangle^2
					     -\langle\sigma_\mathitbf{x}\rangle^2\langle\sigma_{\mathitbf{x}+\mathitbf{r}}\rangle^2\right]}\,.\nonumber
\end{eqnarray}

The definitions in (\ref{eq:GRLA0}), valid at equilibrium, were used in \cite{janus:14b} in an out-of-equilibrium context, for lattices of size $L=80$. In that work
it had been noticed that the replicon is the only correlator that carries a significant signal.

\begin{figure}
 \includegraphics[width=\textwidth]{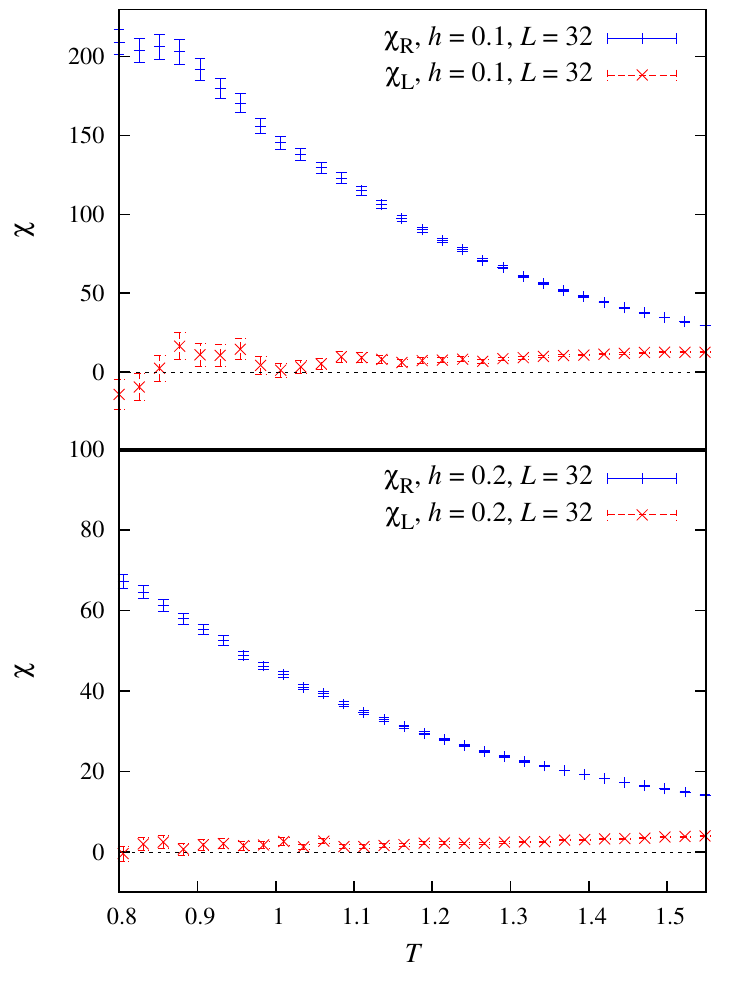}
  \caption{Replicon and longitudinal susceptibility as a function of $T$ in our equilibrium simulations, for the fields $h=0.1,0.2$ 
in our largest lattice sizes ($L=32$).
Just as in \cite{janus:14b} the signal carried by the longitudinal propagator is much smaller than that of
the replicon.
}
 \label{fig:GRGL}
\end{figure}

Also in the present work we measured both signals, and we can confirm
that the same phenomenology is observed in completely thermalised
systems. In figure~\ref{fig:GRGL} we plot both the replicon
susceptibility $\chi_\mathrm{R}$ and the longitudinal susceptibility
$\chi_\mathrm{L}$, at $h=0.1, 0.2$. The figure is qualitatively very
similar to figure~13 of \cite{janus:14b}, where it is shown that
$\chi_\mathrm{R}$ carries a significant signal, while
$\chi_\mathrm{L}$ is very close to zero.\\

\tocless\subsection{The effective anomalous dimension in the spin-glass phase\label{app:eta-eff}}
We can use the fact that $G_\mathrm{R}$ is dominant with respect to $G_\mathrm{L}$ to
predict the value of the effective anomalous exponent $\eta_\mathrm{eff}$ defined in section~\ref{sec:zero-field}
in the deep spin-glass phase.

In fact, in a replica-symmetry breaking (RSB) situation the overlap $q$ has a finite support, so the
overlap's variance $\sigma^2_q = E(q^2)-E(q)^2$ is of order one:
\begin{equation}
 \mathrm{RSB}\Rightarrow \sigma^2_q\sim1\,.
\end{equation}
Now, on general grounds (see for instance \cite{fisher:91}) we can expect
\begin{equation}
\left[E(q^2)-E(q)^2\right] \sim \overline{\langle q^2\rangle-\langle q\rangle^2}\,,
\label{eq:gg}
\end{equation}
and remark that the r.h.s. is $\hat \Gamma_2 ({\bf0})/V$, the 
zero-moment Fourier transform of $\Gamma_2$ [defined in (\ref{eq:gamma2})].
We have then that in RSB conditions
\begin{equation}
 \Gamma_2({\bf0})\sim V\sigma^2_q \stackrel{\mathrm{RSB}}{\sim} V\,.
\end{equation}
$\Gamma_2$ can be related to the replicon and longitudinal susceptibilities through (\ref{eq:gamma2}),
that imply that $\Gamma_2({\bf0}) = \chi_\mathrm{R} + \frac{1}{2}\chi_\mathrm{L}$.
Now, in the beginning of this section we found out empirically that the longitudinal
susceptibility is subdominant with respect to the replicon channel (figure~\ref{fig:GRGL}), so 
in the large-volume limit, in the presence of RSB, the replicon susceptibility scales like the volume:
\begin{equation}
\label{eq:rsb-implies}
 \mathrm{RSB}\Rightarrow\chi_R\sim V\,.
\end{equation}
Let us recall (\ref{eq:quotients}) and impose the just-found implication.
We have then
\begin{equation}
 2^{D} \stackrel{\mathrm{RSB}}{=} \frac{\chi_{\mathrm{R},2L}}{\chi_{\mathrm{R},L}} \equiv 2^{2-\eta_\mathrm{eff}}\,,
\end{equation}
therefore in the spin-glass phase we would have $\eta_\mathrm{eff}=-1$.\\

\tocless\section{Technical details on the creation of quantiles\label{app:hack}}
\addcontentsline{toc}{section}{Appendix C. Technical details on the creation of quantiles}
To grant the reproducibility of our results, we give some details on how we
proceeded in the labelling of the observables with the conditioning variate, and
over the definition of the quantiles.\\

\tocless\subsection{Creating the \texorpdfstring{$P(\hat q)$}{P(q)}}
As already explained in section~\ref{sec:model-simulations}
the analysis we conduct uses instantaneous realisations of the
observables, instead of the average over the equilibrium regime. 
This is because computing $P(\hat q)$ properly requires as many instances of the
overlap as possible.

Operatively, we divide the second half of the simulation time-series in 16
blocks, and for the 4 replicas we save the final configuration of each
block. This gives us $16^4$ configurations over which we can potentially
compute overlaps for a single sample. For $N_\mathrm{t}$ times we pick 4 random numbers
between 1 and 16 to create an instant measure. This way we increase our
statistics of a factor $N_\mathrm{t}$, obtaining ${\cal N_\mathrm{m}}=N_\mathrm{samples}(L,T,h)\times N_\mathrm{t}$
measures for each triplet $(L,T,h)$. We used $N_\mathrm{t}=1000$.

With the 4 replicas it is possible to compute 6 different overlaps $q_i$
($i=1,...,6$), and one instance of most observables, for example the replicon
susceptibility $\chi_\mathrm{R}$. Our ansatz is that $\chi_\mathrm{R}$ and the overlaps have some
type of correlation, so we label $\chi_\mathrm{R}$ with some function of the overlaps
$\hat q(q_1,...,q_6)$, that we called conditioning variate.

The random variable $\hat q$ will have a probability distribution
function $P(\hat q)$ that we want to calculate numerically, in order
to be able to work on the quantiles.  Since our objective is not to
individuate exactly the quantiles, but to compute observables related
to a particular quantile, we coarse grain the range of definition of
the $P(\hat q)$.  This is done by making a binning of the $P(\hat q)$.
This way, each conditioned expectation value of a generic observable,
$E({\cal O}| \hat q)$, can be calculated over a reasonable amount of
measurements, and we have exactly one conditioned expectation value
for each bin of the $P(\hat q)$. Integrals such as those in
(\ref{eq:E-conditioned}) and (\ref{eq:regla-de-suma-var}) are
computed as sums over the histogram bins.  Furthermore, the described
histogramming procedure has the advantage that errors can be
calculated in a very natural way with the jackknife method.

In order to have, as $L$
increases, both a growing number of bins, and of points per bin, we choose bins
of width $\Delta \hat q = 1/\sqrt{aV}$. We add the restriction of having at
least 150 bins, in order to be able to define the quantiles properly (we want
to avoid the eventuality of two quantiles between the same two bins).  We
verified that there is no appreciable difference in the results between
$a=1,2,4$. Larger $a$ implies a too large error, because the bins are too
small, while with smaller $a$ the bins are too few. The results we show
throughout this article have $a=2$.

To compute the conditional expectation values defined in section~\ref{sec:conditional} 
we use the following estimators:
\begin{eqnarray}
 E({\cal O}|\hat q = c)&\approx& \frac{\frac{1}{{\cal N}_\mathrm{m}} \sum_i^{\cal N_\mathrm{m}} {\cal O}_i {\cal X}_{c}(\hat q_i)}
			       {\frac{1}{{\cal N}_\mathrm{m}} \sum_i^{\cal N_\mathrm{m}}{\cal X}_{c}(\hat q_i)}~,\\[2ex]
 P(\hat q)             &\approx&
		\frac{1}{\cal N_\mathrm{m}}\sum_i^{\cal N_\mathrm{m}} {\cal X}_{c=\hat q}(\hat q_i)\,,
\end{eqnarray}
where with the symbol ``$\approx$'' we stress that the quantity is an estimator that converges
to the exact value only in the limit of an infinite number of measurements.\\

\tocless\subsection{Defining the quantiles}
As stated in section~\ref{sec:median}, the quantiles are the points that
separate definite areas under $P(\hat q)$. Therefore, the
$i^\mathrm{th}$ quantile $\tilde q_i$ is defined by means of the
cumulative distribution $X(\hat q)$ of $P(\hat q)$, via the implicit
relation
\begin{equation} 
X(\tilde q_i) = \int_{-1}^{\tilde
q_i} \mathrm{d}\hat q\ P(\hat q) = \frac{i}{10}\,.  
\end{equation} 
Since this is a
continuous relation, and our binning is discrete, it is most probable that
the quantile fall between two neighbouring bins. To evaluate the observables
right at the position of the quantile, we make linear interpolations between
the two bins.

Let us call $i^-_\mathrm{bin}$($i^+_\mathrm{bin}$) the bin just under (over)
quantile $i$. Observable ${\cal O}_i$ at quantile $i$ will be a linear
combination of the values it assumes at $i^-_\mathrm{bin}$ and
$i^+_\mathrm{bin}$:
\begin{equation}
{\cal O}_i = p\, {\cal O}_{i^-_\mathrm{bin}} + (1-p)\, {\cal O}_{i^+_\mathrm{bin}}\,,
\end{equation}
where the interpretation of the indices is straightforward, and $0\leq p\leq1$
is the interpolation weight 
\begin{equation}
  p = \frac{X(\tilde q_i) - X(\hat q_{i^+_\mathrm{bin}})}{X(\hat q_{i^-_\mathrm{bin}}) - X(\hat q_{i^+_\mathrm{bin}})}\,.
\end{equation}

\tocless\section{A caveat for the quantile description\label{app:Pq}}
\addcontentsline{toc}{section}{Appendix D. A caveat for the quantile description}
In the absence of an applied field, the overlap probability distribution function
$P(q)$ is symmetric, with a single peak centred in $q=0$. 
In the presence of a
field, instead, we expect the $P(q)$ to be strictly positive, at least in the
thermodynamic limit. Similarly, we expect that the probability distribution
function $P(q_\mathrm{med})$ have only one peak at positive $q_\mathrm{med}$
when a field is applied, and a peak in $q=0$ if $h=0$.

\begin{figure}[t]
 \centering
 \includegraphics[width=0.8\columnwidth]{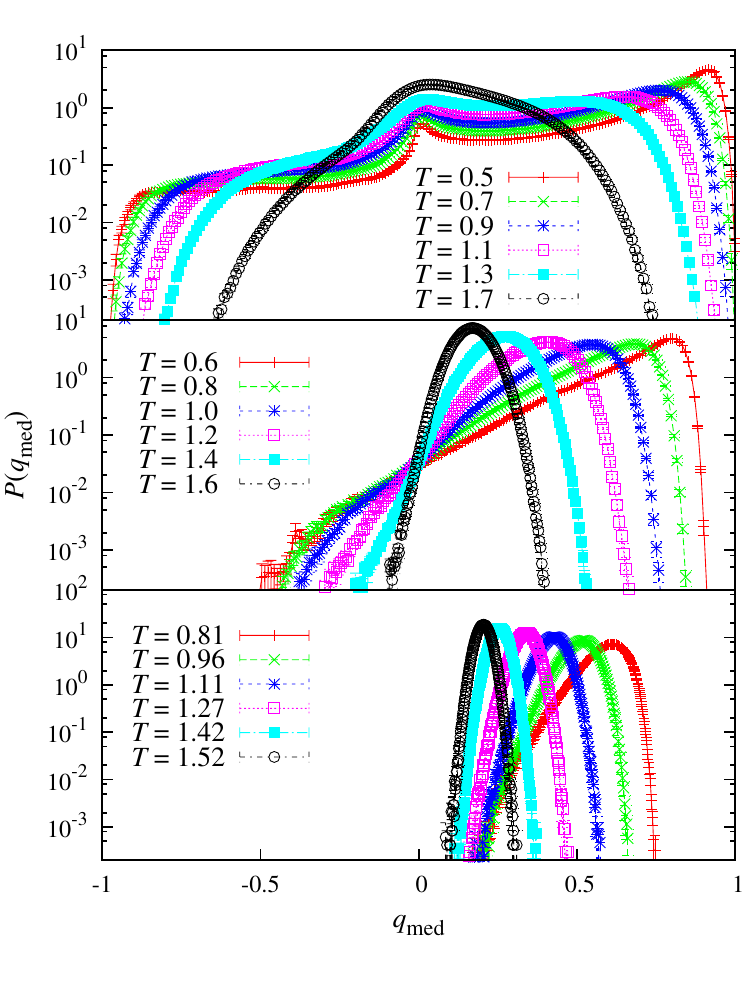}
  \caption{Median overlap probability distribution function $P(q_\mathrm{med})$ with $h=0.2$ 
for different temperatures
(the ones from $L=32$ are an approximation to the second decimal digit).
The top figure shows the case of
 $L=6$, where the lowest temperature curves display a second peak around
 $q_\mathrm{med}\simeq0$, which disappears when $T$ increases.  For $L=16$, the
 $P(q_\mathrm{med})$ are single-peaked, but assume also negative values.  In the
 bottom curve we have $L=32$, where the $P(q_\mathrm{med})$ are single-peaked
and defined only on positive $q_\mathrm{med}$, since we are closest to the
asymptotic behavior. }
\label{fig:PqT}
\end{figure}

If the system sizes are too small, it may occur that the $h=0$ behavior bias the
$P(q_\mathrm{med})$. This is what happens, for example, when $L=6$, $h=0.2$ and
the temperature is sufficiently low: a second peak around
$q_\mathrm{med}\simeq0$ develops upon lowering $T$ (figure~\ref{fig:PqT}, top). 
This second peak disappears when we increase the lattice size (figure~\ref{fig:PqT}, 
centre), and the $P(q_\mathrm{med})$ assumes only positive values when $L$ is large 
enough (figure~\ref{fig:PqT}, bottom).  The lower the field, the easier it is to find
multiple peaks, and the greater the system has to be to be able to neglect the
$h=0$ behavior. For $h=0.05$, even lattices with $L=12$ show a double peak.

A second peak in $P(q_\mathrm{med})$ is a clear signal that we are observing 
and echo of $h=0$. When we make the quantile classification, and have a
quantile on a peak, we are seeing \emph{only} non-asymptotic data. Thus,
quantile 1 for the smallest lattices gives us no relevant information. 

If we plot any observable ${\cal O}$ related to quantile 1 versus the
temperature, the information will be biased for low temperatures, and
the bias will gradually disappear as we increase $T$. The result is
that the curve ${\cal O}(T)$ will have a strange shape and will be of
no use (see, e.g., the $h=0.05$ data in figure~\ref{fig:qL}). This is
why we did not include the $L=6$ points in
figure~\ref{fig:xiL-separil_h02}.\\

\tocless\section{Finding a privileged \texorpdfstring{\boldmath $q$}{q}\label{app:q_L}}
\begin{figure}[tbhp]
\centering
\includegraphics[width=0.85\columnwidth]{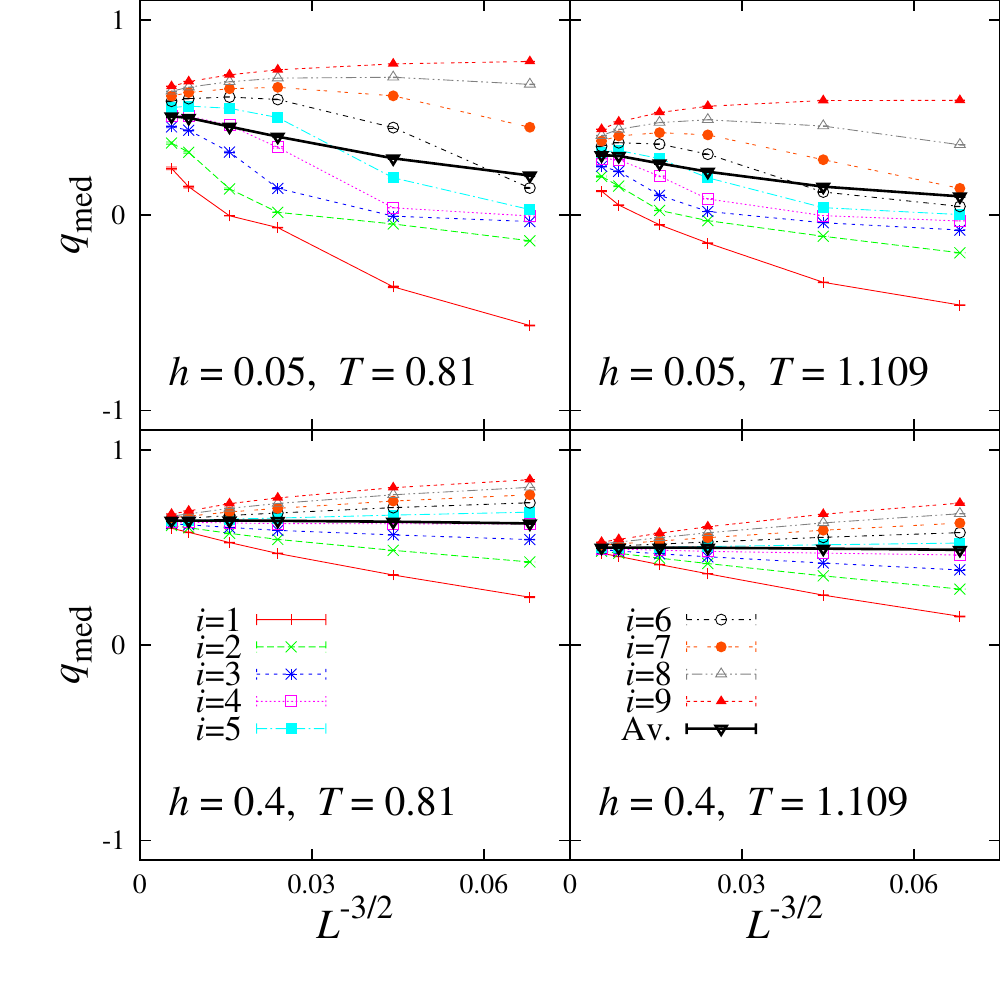}
\caption{Extrapolations to infinite size of the quantile overlap $\tilde q_i$,
for $T=0.81$ (left) and $T=1.109$ (right), and fields $h=0.05$ (top) and $h=0.4$ (bottom). We show
quantiles $i=1,..9$ (thin lines), and the average behavior (bold line). 
The $h=0.4$ extrapolations to 
infinite volume were clean ($\chi^2/\mathrm{DOF} < 1$), while for $h=0.05$ (and all the other fields we simulated),
we encountered too strong and nonlinear finite-size effects to get reasonable extrapolations.
We choose $1/L^{D/2}$ as scaling variable because in conditions of validity of
the central limit theorem, the fluctuations should be of order $1/\sqrt{V}$.}
\label{fig:qL}
\end{figure}

\addcontentsline{toc}{section}{Appendix E. Finding a privileged \texorpdfstring{\boldmath $q$}{q}}
Since all our simulations are in the paramagnetic phase the
thermodynamic limit of the $P(q)$ is a delta function, so all the
quantiles should tend to the a common $q=q_\mathrm{EA}$ in the
$L\rightarrow\infty$ limit.  We tried to perform these extrapolations
at fixed (reasonably low) temperature, to see if we could look at the
problem from such a privileged position.  In figure~\ref{fig:qL} we see
this type of extrapolation for $h=0.4$ and $h=0.05$, at temperatures
$T=0.81$ and $1.109$. The first is the lowest temperature we simulated in
all our lattices, while the second is the zero-field critical
temperature \cite{janus:13b}.  Since we are in the paramagnetic phase
and we are plotting $\tilde q_i$ versus the inverse lattice size, the
curves should cross at the intercept. This is indeed what appears to
happen, but although in the case of $h=0.4$, the extrapolations were
clean, for all the other simulated fields the finite-size effects were
too strong and nonlinear to make solid extrapolations. We remark, yet,
that once $L>8$ the $5^\mathrm{th}$ quantile is the one with the least
finite-size effects.\\

\tocless\section{Quantiles with 2-replica correlators\label{app:2rep}}
\addcontentsline{toc}{section}{Appendix F. Quantiles with 2-replica correlators}
To have well behaving (connected) correlators in the presence of a magnetic field we needed to use
4 replicas for each instance of them. As explained in sections \ref{sec:conditional} and 
\ref{sec:median}, since the overlap is a 2-replica observable, we had to choose a function 
of the 6 overlaps in order to have a one-to-one correspondence between conditioning variates
and the correlators. The functions we tried out were the minimum, the maximum, the median
and the average of the 6 overlaps.

Now, it is legitimate to ask oneself if the fluctuations we observed would also be visible 
having $q$ as conditioning variate. Although this is not possible with the replicon
correlation function $G_\mathrm{R}$, we can renounce to have a connected correlation function,
and study the fluctuations of the 2-replica point-to-plane correlator\\
\begin{equation}\label{eq:g_2_r}
 G_2^\mathrm{nc}(r) = \sum_{y,z} E(\,q_{(0,0,0)}\, q_{(r,y,z)}\,) \,,
\end{equation}
which allows us to have $q$ as a conditioning
variate. $G_2^\mathrm{nc}(r)$ is the total correlation between the
origin, $(0,0,0)$, and the plane $x=r$. Of course, one could
equivalently consider the planes $y=r$ or $z=r$. One can displace
freely the origin, as well. We average over all these $3V$ choices.

\begin{figure}[t]
 \includegraphics[width=\columnwidth]{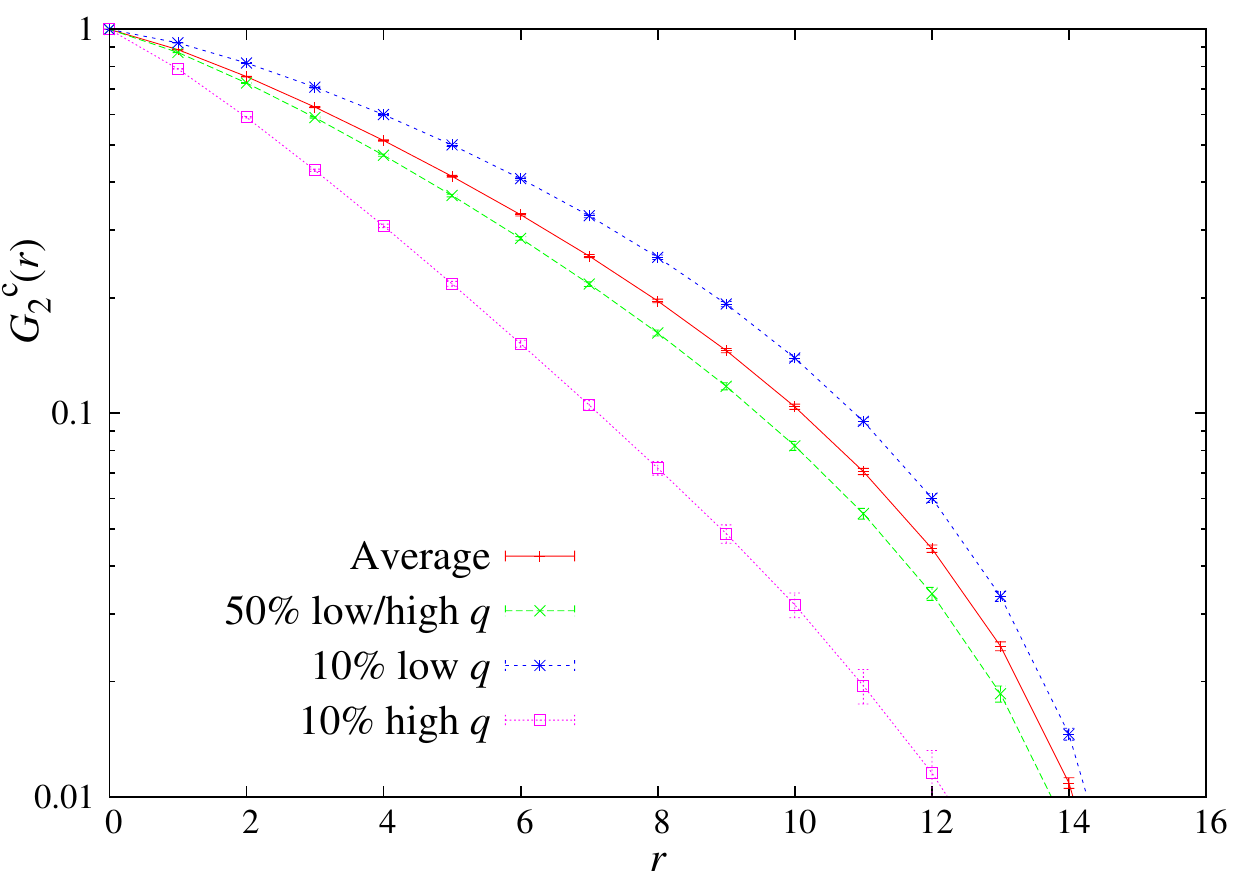}
 \caption{Same as figure~\ref{fig:Cr}, but for the 2-replica connected
   correlation function $G_2^\mathrm{c}(r)$ (\ref{eq:g_2_r_norm}). We show $L=32$ data from $h=0.2$,
   $T=0.805128$.  Note that $G_2^\mathrm{c}(r)$ is bound to be 1 at
   $r=0$, and 0 at $r=L/2$, so the fluctuations between different
   quantiles are even stronger than they may appear.}
\label{fig:G2}
\end{figure}

At this point, it is possible to compare with previous work that
studied fluctuations with 2-replica correlators \cite{parisi:12b}.
Furthermore, we can construct the pseudoconnected correlation function
\begin{equation}\label{eq:g_2_r_norm}
 G_2^\mathrm{c}(r) = \frac{G_2^\mathrm{nc}(r) - G_2^\mathrm{nc}(L/2)}{G_2^\mathrm{nc}(0) - G_2^\mathrm{nc}(L/2)} \,,
\end{equation}
which forcedly is one for $r=0$, and goes to zero for $r=L/2$. In figure~\ref{fig:G2} we show that
the same dramatic fluctuations encountered with $G_\mathrm{R}$ (figure \ref{fig:Cr}) are also present here.

\begin{figure}[tbh]
 \includegraphics[width=0.495\columnwidth]{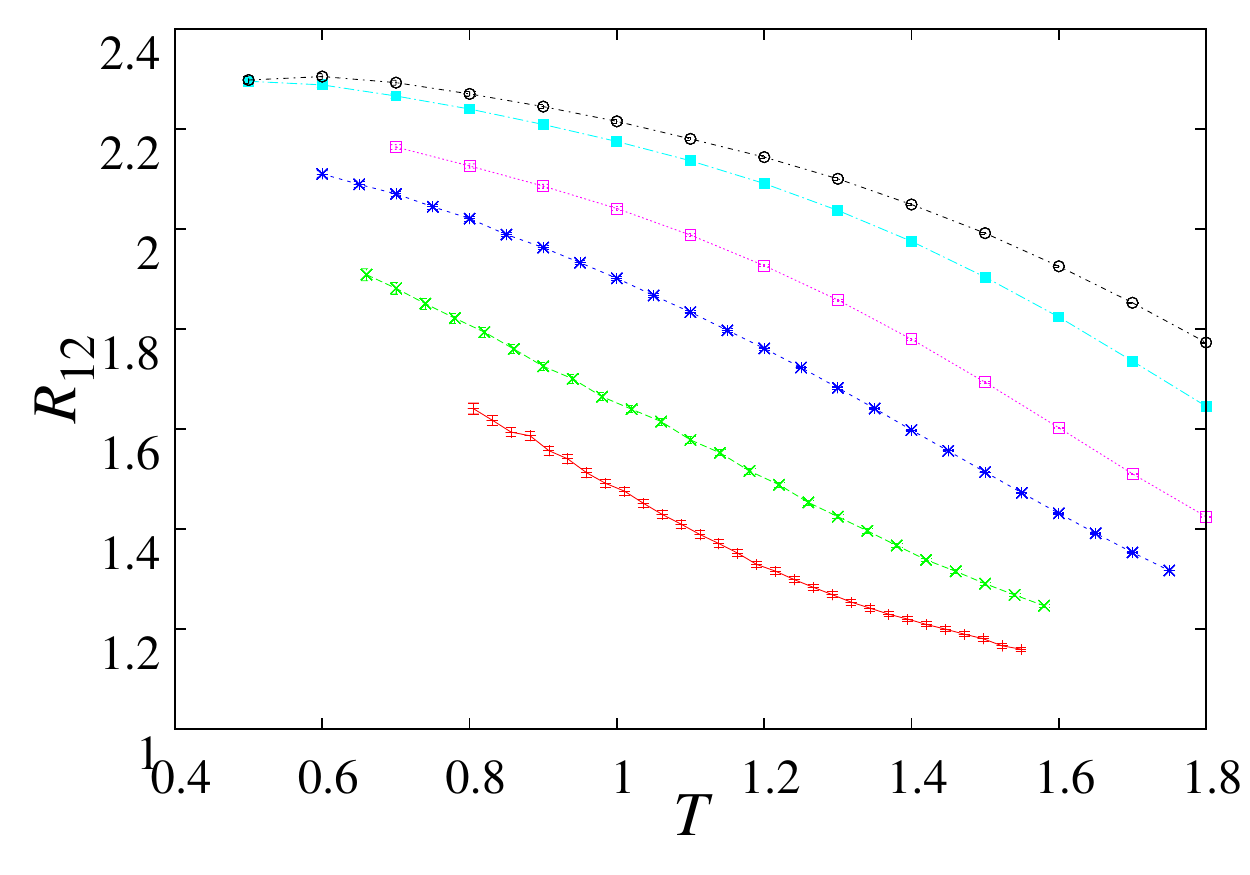}
 \includegraphics[width=0.495\columnwidth]{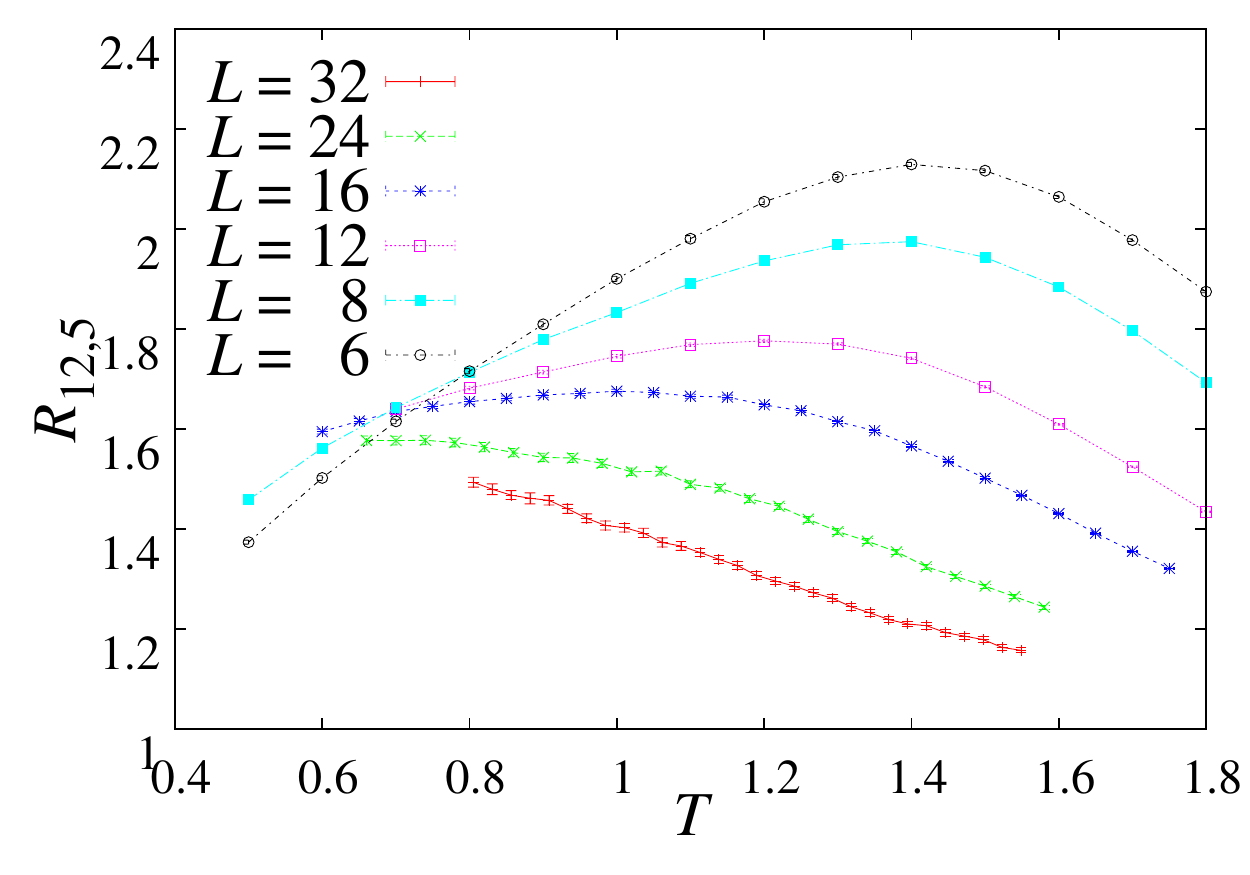}
 \caption{The $R_{12}$ cumulant computed from the two-replica
   correlation function (\ref{eq:g_2_r}) rather than from four
   replicas. The field is $h=0.2$.  On the left side we show the
   average behavior, and on the right, the $5^\mathrm{th}$ quantile,
   with the plain overlap $q$ (\ref{eq:q}) as conditioning
   variate.}
\label{fig:R12-global-mediana_2rep}
\end{figure}

The overall results, figure~\ref{fig:R12-global-mediana_2rep}, are
consistent with the picture we draw in the main part of the paper. On
the one hand, the standard data average hides all signs of a phase
transition.  On the other hand, the fifth quantile displays signs of
scale invariance.

\tocless\section{High-temperature extrapolation of the critical line\label{app:HT}}
\addcontentsline{toc}{section}{Appendix G. High-temperature extrapolations}

\begin{figure}[tbh]
\centering
 \includegraphics[height=0.7\columnwidth,angle=270]{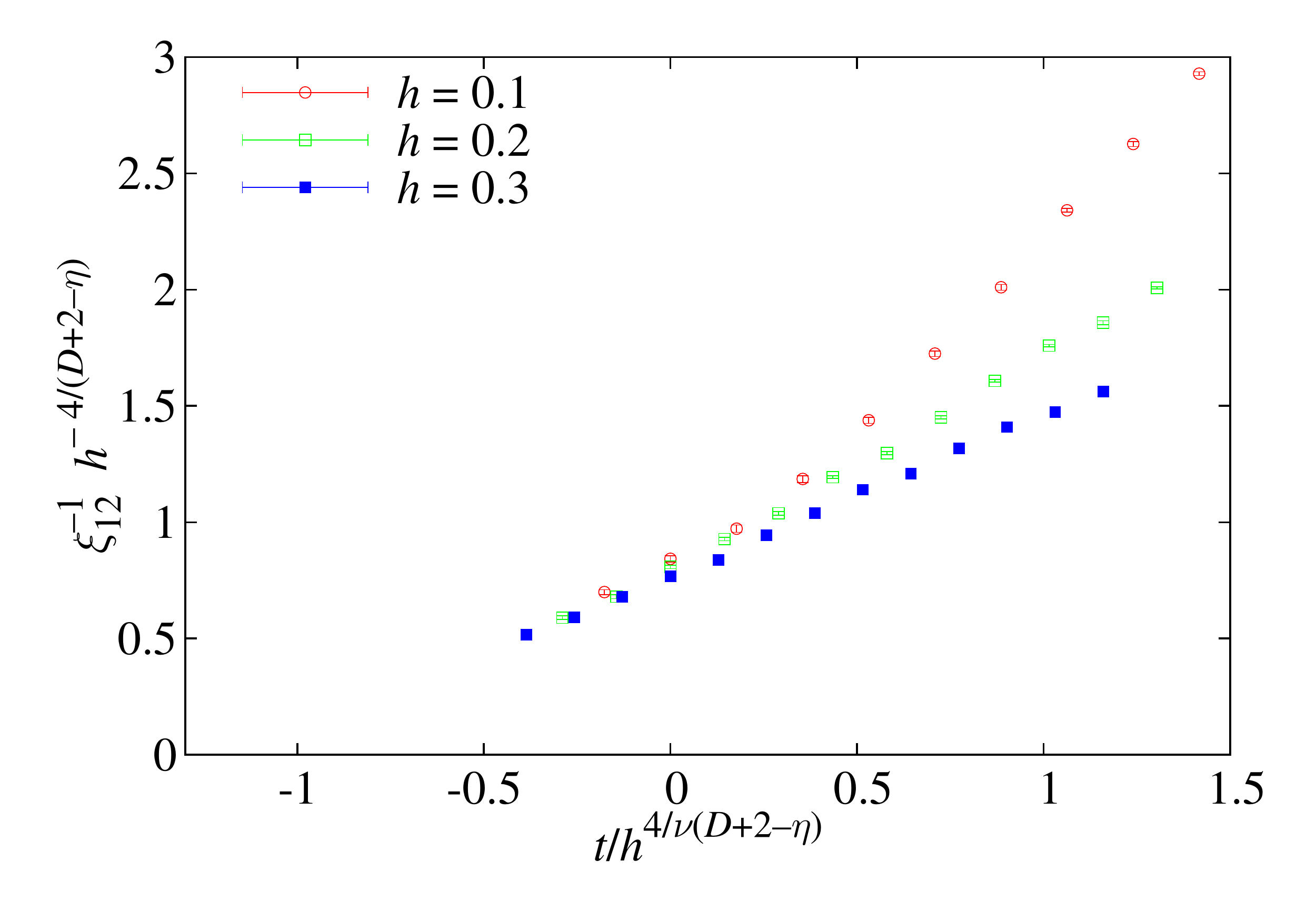}
 \caption{The  $\xi_{12}$ estimate of the correlation length, from the
thermodynamic-limit data in \cite{janus:14b}. We display the data as
suggested by the Fisher-Sompolinsky scaling (\ref{eq:xi-FS}). We
plot $1/\xi$, rather than $\xi$, in order to identify a possible critical
point.
}
\label{fig:xi-FS}
\end{figure}

While this paper was being completed, M. A. Moore and J. Yeo pointed out
to us that it would be worthy to consider the Fisher-Sompolinsky
scaling~\cite{fisher:85} for the correlation-length data from
\cite{janus:14b}. Since these data were obtained in a very large
lattice at comparatively high temperatures, the large fluctuations
discussed in this paper should be much attenuated. Therefore, the
results in \cite{janus:14b} are to be regarded as representative of
the thermodynamic limit in the paramagnetic phase.

The Fisher-Sompolinsky scaling suggests a possible location of the dAT
line in rough agreement with our finite-size scaling analysis at lower
temperatures. Let us see how it works.

A standard RG argument (see, e.g.,~\cite{amit:05}) tells us that, at
least for small magnetic fields and close to the zero-field critical
temperature, the correlation length should behave as
\begin{equation}\label{eq:xi-FS}
h ^{2/y_h} \xi(t,h)= F\big(t/h^{2y_t/y_h}\big)\,,
\end{equation}
where $t$ is the $h=0$ reduced temperature, $t=(T-T_\mathrm{c})/T_\mathrm{c}$,
while the $h=0$ critical exponents are
\begin{equation}
y_h=\frac{D+2-\eta}{2}\,,\quad y_t=\frac{1}{\nu}\,.
\end{equation}
We shall assess (\ref{eq:xi-FS}) using the estimates of
\cite{janus:13b}.\footnote{The correlation length in \cite{janus:14b} is
not exactly the same computed here (technically, it is the so-called
$\xi_{12}$ integral estimator~\cite{janus:09b}). Yet, the difference
is immaterial as far as scaling properties are concerned.}

In figure~\ref{fig:xi-FS} we plot $y=1/[h ^{2/y_h} \xi(t,h)]$ as a
function of the scaling variable $x=t/h^{2y_t/y_h}$.  At the
(would-be) dAT line, $y$ should vanish at a critical $x_c$. In fact,
$y$ should behave as $y\sim (x_\mathrm{c}-x)^{\nu_h}$ (i.e., $\nu_h$ is
the correlation-length exponent in a field). Hence, at least for small
$h$, one would have
\begin{equation}
T_\mathrm{c}(h)=T_\mathrm{c}(h=0) \big[1+x_\mathrm{c} h^{2y_t/y_h}\big]
\end{equation}

Now, the scaling in figure~\ref{fig:xi-FS} is poor, but it improves upon
decreasing $x$ (i.e., lowering the temperature). The estimation of
$x_\mathrm{c}$ depends strongly on the value that one uses for exponent
$\nu_h$. For instance, the rather improbable value $\nu_h=1$ suggests
$x_\mathrm{c}\approx -1.2$. With a rather more plausible
$\nu_h=\nu\approx 2.56$ one would guess $x_\mathrm{c}\approx -0.6$.
This disparate range of $x_c$ would predict
$T_\mathrm{c}(h=0.1)\approx$ 0.4--0.8, $T_\mathrm{c}(h=0.2) \approx$ 0.3--0.7, $T_\mathrm{c}(h=0.4)\approx$ 0.1--0.6.
Overall, these numbers are not in conflict with the finite-size scaling
analysis we presented, but they are not of much help to locate of the dAT line.

\section*{References}
\addcontentsline{toc}{section}{References}

\bibliography{/homenfs/rg/biblio}
\bibliographystyle{iopart-num}

\end{document}